\documentclass[reqno,12pt,a4paper]{amsart}

\usepackage{mathrsfs}
\usepackage{amssymb}
\usepackage{amsfonts}
\usepackage{latexsym}
\usepackage{amsthm}

\usepackage{graphicx}
\def\lb{\label}

\newcommand{\er}[1]{\textrm{(\ref{#1})}}

\begin{document}


\renewcommand{\theequation}{\arabic{section}.\arabic{equation}}
\theoremstyle{plain}
\newtheorem{theorem}{\bf Theorem}[section]
\newtheorem{lemma}[theorem]{\bf Lemma}
\newtheorem{corollary}[theorem]{\bf Corollary}
\newtheorem{proposition}[theorem]{\bf Proposition}
\newtheorem{definition}[theorem]{\bf Definition}
\newtheorem{remark}[theorem]{\it Remark}

\def\a{\alpha}  \def\cA{{\mathcal A}}     \def\bA{{\bf A}}  \def\mA{{\mathscr A}}
\def\b{\beta}   \def\cB{{\mathcal B}}     \def\bB{{\bf B}}  \def\mB{{\mathscr B}}
\def\g{\gamma}  \def\cC{{\mathcal C}}     \def\bC{{\bf C}}  \def\mC{{\mathscr C}}
\def\G{\Gamma}  \def\cD{{\mathcal D}}     \def\bD{{\bf D}}  \def\mD{{\mathscr D}}
\def\d{\delta}  \def\cE{{\mathcal E}}     \def\bE{{\bf E}}  \def\mE{{\mathscr E}}
\def\D{\Delta}  \def\cF{{\mathcal F}}     \def\bF{{\bf F}}  \def\mF{{\mathscr F}}
\def\c{\chi}    \def\cG{{\mathcal G}}     \def\bG{{\bf G}}  \def\mG{{\mathscr G}}
\def\z{\zeta}   \def\cH{{\mathcal H}}     \def\bH{{\bf H}}  \def\mH{{\mathscr H}}
\def\e{\eta}    \def\cI{{\mathcal I}}     \def\bI{{\bf I}}  \def\mI{{\mathscr I}}
\def\p{\psi}    \def\cJ{{\mathcal J}}     \def\bJ{{\bf J}}  \def\mJ{{\mathscr J}}
\def\vT{\Theta} \def\cK{{\mathcal K}}     \def\bK{{\bf K}}  \def\mK{{\mathscr K}}
\def\k{\kappa}  \def\cL{{\mathcal L}}     \def\bL{{\bf L}}  \def\mL{{\mathscr L}}
\def\l{\lambda} \def\cM{{\mathcal M}}     \def\bM{{\bf M}}  \def\mM{{\mathscr M}}
\def\L{\Lambda} \def\cN{{\mathcal N}}     \def\bN{{\bf N}}  \def\mN{{\mathscr N}}
\def\m{\mu}     \def\cO{{\mathcal O}}     \def\bO{{\bf O}}  \def\mO{{\mathscr O}}
\def\n{\nu}     \def\cP{{\mathcal P}}     \def\bP{{\bf P}}  \def\mP{{\mathscr P}}
\def\r{\rho}    \def\cQ{{\mathcal Q}}     \def\bQ{{\bf Q}}  \def\mQ{{\mathscr Q}}
\def\s{\sigma}  \def\cR{{\mathcal R}}     \def\bR{{\bf R}}  \def\mR{{\mathscr R}}
\def\S{\Sigma}  \def\cS{{\mathcal S}}     \def\bS{{\bf S}}  \def\mS{{\mathscr S}}
\def\t{\tau}    \def\cT{{\mathcal T}}     \def\bT{{\bf T}}  \def\mT{{\mathscr T}}
\def\f{\phi}    \def\cU{{\mathcal U}}     \def\bU{{\bf U}}  \def\mU{{\mathscr U}}
\def\F{\Phi}    \def\cV{{\mathcal V}}     \def\bV{{\bf V}}  \def\mV{{\mathscr V}}
\def\P{\Psi}    \def\cW{{\mathcal W}}     \def\bW{{\bf W}}  \def\mW{{\mathscr W}}
\def\o{\omega}  \def\cX{{\mathcal X}}     \def\bX{{\bf X}}  \def\mX{{\mathscr X}}
\def\x{\xi}     \def\cY{{\mathcal Y}}     \def\bY{{\bf Y}}  \def\mY{{\mathscr Y}}
\def\X{\Xi}     \def\cZ{{\mathcal Z}}     \def\bZ{{\bf Z}}  \def\mZ{{\mathscr Z}}
\def\O{\Omega}

\newcommand{\mc}{\mathscr {c}}

\newcommand{\gA}{\mathfrak{A}}          \newcommand{\ga}{\mathfrak{a}}
\newcommand{\gB}{\mathfrak{B}}          \newcommand{\gb}{\mathfrak{b}}
\newcommand{\gC}{\mathfrak{C}}          \newcommand{\gc}{\mathfrak{c}}
\newcommand{\gD}{\mathfrak{D}}          \newcommand{\gd}{\mathfrak{d}}
\newcommand{\gE}{\mathfrak{E}}
\newcommand{\gF}{\mathfrak{F}}           \newcommand{\gf}{\mathfrak{f}}
\newcommand{\gG}{\mathfrak{G}}           
\newcommand{\gH}{\mathfrak{H}}           \newcommand{\gh}{\mathfrak{h}}
\newcommand{\gI}{\mathfrak{I}}           \newcommand{\gi}{\mathfrak{i}}
\newcommand{\gJ}{\mathfrak{J}}           \newcommand{\gj}{\mathfrak{j}}
\newcommand{\gK}{\mathfrak{K}}            \newcommand{\gk}{\mathfrak{k}}
\newcommand{\gL}{\mathfrak{L}}            \newcommand{\gl}{\mathfrak{l}}
\newcommand{\gM}{\mathfrak{M}}            \newcommand{\gm}{\mathfrak{m}}
\newcommand{\gN}{\mathfrak{N}}            \newcommand{\gn}{\mathfrak{n}}
\newcommand{\gO}{\mathfrak{O}}
\newcommand{\gP}{\mathfrak{P}}             \newcommand{\gp}{\mathfrak{p}}
\newcommand{\gQ}{\mathfrak{Q}}             \newcommand{\gq}{\mathfrak{q}}
\newcommand{\gR}{\mathfrak{R}}             \newcommand{\gr}{\mathfrak{r}}
\newcommand{\gS}{\mathfrak{S}}              \newcommand{\gs}{\mathfrak{s}}
\newcommand{\gT}{\mathfrak{T}}             \newcommand{\gt}{\mathfrak{t}}
\newcommand{\gU}{\mathfrak{U}}             \newcommand{\gu}{\mathfrak{u}}
\newcommand{\gV}{\mathfrak{V}}             \newcommand{\gv}{\mathfrak{v}}
\newcommand{\gW}{\mathfrak{W}}             \newcommand{\gw}{\mathfrak{w}}
\newcommand{\gX}{\mathfrak{X}}               \newcommand{\gx}{\mathfrak{x}}
\newcommand{\gY}{\mathfrak{Y}}              \newcommand{\gy}{\mathfrak{y}}
\newcommand{\gZ}{\mathfrak{Z}}             \newcommand{\gz}{\mathfrak{z}}

\def\ve{\varepsilon} \def\vt{\vartheta} \def\vp{\varphi}  \def\vk{\varkappa}
\def\vr{\varrho} \def\vs{\varsigma}

\def\A{{\mathbb A}} \def\B{{\mathbb B}} \def\C{{\mathbb C}}
\def\dD{{\mathbb D}} \def\E{{\mathbb E}} \def\dF{{\mathbb F}} \def\dG{{\mathbb G}}
\def\H{{\mathbb H}}\def\I{{\mathbb I}} \def\J{{\mathbb J}} \def\K{{\mathbb K}}
\def\dL{{\mathbb L}}\def\M{{\mathbb M}} \def\N{{\mathbb N}} \def\dO{{\mathbb O}}
\def\dP{{\mathbb P}} \def\dQ{{\mathbb Q}} \def\R{{\mathbb R}}\def\S{{\mathbb S}} \def\T{{\mathbb T}}
\def\U{{\mathbb U}} \def\V{{\mathbb V}}\def\W{{\mathbb W}} \def\X{{\mathbb X}}
\def\Y{{\mathbb Y}} \def\Z{{\mathbb Z}}

\def\dk{{\Bbbk}}


\def\la{\leftarrow}              \def\ra{\rightarrow}            \def\Ra{\Rightarrow}
\def\ua{\uparrow}                \def\da{\downarrow}
\def\lra{\leftrightarrow}        \def\Lra{\Leftrightarrow}


\def\lt{\biggl}                  \def\rt{\biggr}
\def\ol{\overline}               \def\wt{\widetilde}
\def\no{\noindent}


\let\ge\geqslant                 \let\le\leqslant
\def\lan{\langle}                \def\ran{\rangle}
\def\/{\over}                    \def\iy{\infty}
\def\sm{\setminus}               \def\es{\emptyset}
\def\ss{\subset}               \def\ts{\times}
\def\sse{\subseteq}
\def\pa{\partial}                \def\os{\oplus}
\def\om{\ominus}                 \def\ev{\equiv}
\def\iint{\int\!\!\!\int}        \def\iintt{\mathop{\int\!\!\int\!\!\dots\!\!\int}\limits}
\def\el2{\ell^{\,2}}             \def\1{1\!\!1}
\def\sh{\sharp}
\def\wh{\widehat}
\def\bs{\backslash}
\def\intl{\int\limits}

\def\na{\mathop{\mathrm{\nabla}}\nolimits}
\def\sh{\mathop{\mathrm{sh}}\nolimits}
\def\ch{\mathop{\mathrm{ch}}\nolimits}
\def\where{\mathop{\mathrm{where}}\nolimits}
\def\all{\mathop{\mathrm{all}}\nolimits}
\def\as{\mathop{\mathrm{as}}\nolimits}
\def\Area{\mathop{\mathrm{Area}}\nolimits}
\def\arg{\mathop{\mathrm{arg}}\nolimits}
\def\const{\mathop{\mathrm{const}}\nolimits}
\def\det{\mathop{\mathrm{det}}\nolimits}
\def\diag{\mathop{\mathrm{diag}}\nolimits}
\def\diam{\mathop{\mathrm{diam}}\nolimits}
\def\dim{\mathop{\mathrm{dim}}\nolimits}
\def\dist{\mathop{\mathrm{dist}}\nolimits}
\def\Im{\mathop{\mathrm{Im}}\nolimits}
\def\Iso{\mathop{\mathrm{Iso}}\nolimits}
\def\Ker{\mathop{\mathrm{Ker}}\nolimits}
\def\Lip{\mathop{\mathrm{Lip}}\nolimits}
\def\rank{\mathop{\mathrm{rank}}\limits}
\def\Ran{\mathop{\mathrm{Ran}}\nolimits}
\def\Re{\mathop{\mathrm{Re}}\nolimits}
\def\Res{\mathop{\mathrm{Res}}\nolimits}
\def\res{\mathop{\mathrm{res}}\limits}
\def\sign{\mathop{\mathrm{sign}}\nolimits}
\def\span{\mathop{\mathrm{span}}\nolimits}
\def\supp{\mathop{\mathrm{supp}}\nolimits}
\def\Tr{\mathop{\mathrm{Tr}}\nolimits}
\def\BBox{\hspace{1mm}\vrule height6pt width5.5pt depth0pt \hspace{6pt}}
\def\where{\mathop{\mathrm{where}}\nolimits}
\def\as{\mathop{\mathrm{as}}\nolimits}


\newcommand\nh[2]{\widehat{#1}\vphantom{#1}^{(#2)}}
\def\dia{\diamond}

\def\Oplus{\bigoplus\nolimits}



\def\qqq{\qquad}
\def\qq{\quad}
\let\ge\geqslant
\let\le\leqslant
\let\geq\geqslant
\let\leq\leqslant
\newcommand{\ca}{\begin{cases}}
\newcommand{\ac}{\end{cases}}
\newcommand{\ma}{\begin{pmatrix}}
\newcommand{\am}{\end{pmatrix}}
\renewcommand{\[}{\begin{equation}}
\renewcommand{\]}{\end{equation}}
\def\eq{\begin{equation}}
\def\qe{\end{equation}}
\def\[{\begin{equation}}
\def\bu{\bullet}
\def\ced{\centerdot}
\def\tes{\textstyle}


\title[{Estimates, asymptotics and trace formulas for periodic VNLS equations
 }]
{Estimates, asymptotics and trace formulas  for periodic vector NLS equations, II}

\date{\today}

\author[Evgeny Korotyaev]{Evgeny Korotyaev}
\address{Department of Analysis,
Saint-Petersburg State University,   Universitetskaya nab. 7/9, St.
Petersburg, 199034, Russia, \ korotyaev@gmail.com, \
e.korotyaev@spbu.ru}


\begin{abstract}
\no We consider a first order operator with a smooth periodic 3x3 matrix
potential on the real line. It is the Lax operator for the periodic
vector NLS equation.  Its spectrum covers the real
line and it is union of the spectral bands of multiplicity 3, separated by
intervals (gaps) of multiplicity 1.  We prove and describe the following:
 \\
$\cdot$ The geometry of the Riemann surface and its branch points.
\\
$\cdot$ The asymptotics of  branch points are determined and they
are real at high energy.
\\
$\cdot$  Trace formulas for integral of motions, including the Hamiltonian
of the NLS equation.
\\
$\cdot$ Estimates of the Hamiltonian in terms of gap lengths.
\\
 The proof is based on the analysis of averaged  quasi-momentum as a conformal mapping of the upper half plane on the domain on the upper half plane and on the asymptotics of the monodromy matrix and multipliers at high energy.

\end{abstract}

\maketitle

\begin{quotation}
\begin{center}
{\bf Table of Contents}
\end{center}

\vskip 6pt

{\footnotesize

1. Introduction and main results \hfill \pageref{Sec1}\ \ \ \ \

2. Preliminary results for   perturbations from $L^2$
 \hfill
\pageref{Sec2}\ \ \ \ \

3. The monodromy matrix and its trace for smooth perturbations  \hfill \pageref{Sec3}\ \ \ \ \

4. The Riemann surface and its branch points\hfill \pageref{Sec4}\ \
\ \ \

5. Lyapunov functions \hfill \pageref{Sec5}\ \ \ \ \

6. Constants of motion, conformal mappings and trace formulas \hfill
\pageref{Sec6}\ \ \ \ \

7.  Appendix \hfill \pageref{Sec7}\ \ \ \ \

 }
\end{quotation}

\section {Introduction and main results \lb{Sec1}}
\setcounter{equation}{0}

\subsection{Vector Nonlinear Schr\"odinger equations}
Consider the periodic 2 dimensional defocusing Vector Nonlinear
Schr\"odinger equation (or the Manakov systems, or shortly the MS)
\[
\lb{VSE1}
\begin{aligned}
iv_t=-v_{xx}+ 2|v|^2v,\qqq v=(v_1,v_2)^\top \in\C^2,
\end{aligned}
\]
where $|v|^2=|v_1|^2+|v_2|^2$. Here $v_1, v_2$ represents as a sum
of right- and left-hand polarized waves. The  NLS equation is one of
the most fundamental and the most universal nonlinear PDE. The vector
NLS equation appears in physics to study two mode optical fibers,
photorefractive materials and so on, see \cite{BN67}, \cite{GT73} \cite{SR19}--
\cite{W13}. Manakov \cite{Ma74} obtained
first basis results about the vector case and proved that it is
completely integrable. 
In order to study various invariants of the MS
we consider Manakov  operators $\cL$ (i.e., the Lax operators for the MS)  on the real line given by
\[
\lb{do1} \tes
 \cL=iJ{d\/dx}+V, \qqq J=\diag (1,-1,-1),\qqq  V=\ma 0& v^*\\
v&0\am=V^*,
\]
We assume that the 1-periodic $1\ts 2$ matrix
(vector) $v$ satisfies
\[
\lb{dv}
\begin{aligned}
v=(v_1,v_2)^\top\in \mH=L^2(\T)\os L^2(\T), \qqq \T=\R/\Z, \\
\end{aligned}
\]
where  $\mH$ is the complex Hilbert space equipped with the norm
$\|v\|^2=\!\!\int_0^1 (|v_1|^2+|v_2|^2)dx$.

There are a lot of results devoted to systems with periodic coefficients,
see  \cite{BBK06}, \cite{CK06}, \cite{CG02}, \cite{C00}, \cite{GL55},  \cite{Kr55}, \cite{RS78}, \cite{YS75} and see  references therein.
We mention results relevant for our context. Schr\"odinger operators
with $d\ts d$ matrix valued periodic potentials are discussed in
\cite{CK06}  and the case $d=2$ in \cite{BBK06}. The first order
systems were studied in \cite{K08}, \cite{K10}. In \cite{K10} we
considered the Lax type operators (for the matrix NLS equations), where the
matrix $J=\1_{m}\os -\1_{n}$ for some $m,n\in \N$, the case $m=n$ is
considered in \cite{K08}.

Zakharov and Shabat \cite{ZS72} obtained
first basis results about the scalar NLS equation case and proved that it is
completely integrable.  For the scalar
NLS equation we consider the Zakharov-Shabat (shortly ZS) operators
ZS-operator $\cL_{zs}$ on $L^2(\R,\C^2)$ given by
\[
\lb{ddzs} \tes \cL_{zs}=iJ_{zs}{d\/ dx}+ V_{zs},\ \ \ \  J_{zs} =\ma
1&0\\ 0&-1\am, \ \ \ \ \ V_{zs}= \ma 0 & \ol u\\  u & 0\am,\qq u\in
L^2(\T),
\]
see, e.g., \cite{LS91}. The spectrum of $\cL_{zs}$ is purely
absolutely continuous and consists of spectral bands separated by
the gaps  $\g_n$:
\[
\lb{ddzs1} \s(\cL_{zs})=\R\sm \cup \g_n, \qq
\g_n=(\l_n^-,\l_n^+),\qq \l_n^\pm=\pi n+o(1)\ \ \as \ n\to \pm \iy.
\]
Here $\l_n^\pm, n\in \Z$ are 2-periodic eigenvalues of the equation
$iJ_{zs}y'+ V_{zs}y=\l y$. The corresponding Riemann surface is
two-sheeted with the real branch points $\l_n^\pm, n\in \Z$. There
are a lot of results about inverse problems for ZS-operators, see
\cite{GK14}, \cite{K01}, \cite{K05}, and references therein.

\subsection{Preliminary results}
In order to study various invariants of the MS we 
introduce the $3\ts 3$-matrix valued solutions $y(x,\l), v\in\mH$ of the
Manakov system
\[
\lb{1} \tes iJ y'+V y=\l y,\qqq x\ge 0,\qq  y(0,\l)=\1_{3}\qq \forall \
\l\in\C.
\]
Define the monodromy matrix $\p(\l)=y(1,\l)$ and its determinant
$D(\t,\l)$ by
\[
\lb{2} D(\t,\l)=\det (\p(\l)-\t \1_{3})=-\t^3+\t^2T-\t e^{i\l } \wt
T+e^{i\l},\qqq \t,\l\in \C,
\]
where the coefficients $T(\l)=\Tr \p(\l)$ and $\wt T(\l)=\ol T(\ol
\l)$  are entire functions of $\l$. An eigenvalue $\t$ of $\p(\l)$
is called a {\it multiplier}, it is a root of the algebraic equation
$D(\t,\l)$, where $\t,\l\in\C$. Let $\t_j=\t_j(\l), j=1,2,3$ be multipliers of $\p(\l)$. Due to \er{2} there exist exactly three roots $\t_1(\cdot), \t_2(\cdot)$ and $\t_3(\cdot)$, which constitute three distinct branches of some analytic function $\t(\cdot)$ on some
3-sheeted or 2-sheeted Riemann surface $\cR$ that has only algebraic singularities in any bounded domain from $\C$, see, e.g.,  Ch.~4 in \cite{S57}. If $v=0$, then the corresponding multipliers are given by
$$
\t_1^o(\l)=\t_2^o(\l)=e^{i\l},\qq  {\rm and} \qq \t_3^o(\l)=e^{-i\l}.
$$
 The multipliers $\t_j$  have asymptotics from \cite{K24}
\[
\lb{asm}
  \t_j(\l)=\t_j^o(\l)(1+o(1))\qq \as \qq |\l|\to \iy,
 \qq \forall \  j\in \N_3:=\{1,2,3\},
\]
for  $|\l-\pi n|\ge \d, $ for all $n\in \Z$
 and for some small $\d\in (0,{1\/2})$. In order to describe
spectrum of $\cL$ and the Riemann surface it is convenient to define
the Lyapunov function $\D={1\/2}(\t+\t^{-1})$ on $\cR$, where each
branch of the function $\D$ is given by
\[
\lb{deL} \tes \D_j(\l)={1\/2}(\t_j(\l)+\t_j^{-1}(\l)), \qqq j\in
\N_3.
\]
 The spectrum $\s(\cL)$ of $\cL$ is absolutely continuous   and satisfies
\[
\lb{dsp} \s(\cL)=\s_{ac}(\cL)=\R=\gS_1\cup \gS_3,\qqq\qqq
\gS_2=\es,
\]
see \cite{K24}, where  $\gS_j, j\in \N_3$ is the part of the
spectrum of $\cL$ having the multiplicity $j$:
$$
\gS_3=\Big\{\l\in\R:|\t_j(\l)|=1,\ \forall\  j\in \N_3\Big\},\
\gS_1=\Big\{\l\in\R:|\t_j(\l)|=1,\ \text{only \ one }\ j\in
\N_3\Big\},
$$
and $\gS_2$ has the spectrum of $\cL$ with the multiplicity $2$.
In order to describe these spectrum  and the branch points we introduce the {\it discriminant}
$\gD(\l)$ of the polynomial $D(\cdot,\l)$ by
\[
\lb{rd} 
\tes
\gD(\l)=-{e^{-2i\l}\/64}(\t_1-\t_2)^2(\t_1-\t_3)^2(\t_2-\t_3)^2,\qq \l\in\C.
\]
In fact, we have introduced the modified discriminant, since due to
the additional factor $e^{-2i\l}$ our function $\gD$ is real on the
real line. A zero of $\gD$ is a branch point of the corresponding
Riemann surface $\cR$ and the real ones are the end points of th spectral interval, since  $\gD$ satisfies
\[
\lb{sro} \gS_1= \{\l\in\R: \gD(\l)<0\}, \qqq \gS_3= \{\l\in\R:
\gD(\l)\in [0,1]\}.
\]
Recall results about
2-sheeted Riemann surfaces from \cite{K24}, see more in Sect. 2.

 \begin{theorem}
\lb{T1}  The surface $\cR$ is  2-sheeted iff $v=ue$, where $(u,e)\in
L^2(\T)\ts \C^2$ ,  $|e|=1$.

\end{theorem}

2-sheeted Riemann surfaces are well understood and we describe  2-sheeted Riemann surfaces in Sect. 2.  Thus below we consider only 3-sheeted Riemann surfaces.

\subsection{Main results}

The matrix $\mV=\int_0^1V_x^2dx$ has eigenvalues $\gb_j\ge 0, j\in \N_3$
given by:
\[
\lb{V23xx}
\begin{aligned}\tes
 \gb_3=\|v\|^2,\qq \gb_2={\|v\|^2+ \sqrt{\gb_o}\/2}, \qq
\gb_1={\|v\|^2- \sqrt\gb_o\/2},\qq
 \gb_o=(\|v_1\|^2-\|v_2\|^2)^2+4|\gc_{12}|^2,
\end{aligned}
\]
where $\gc_{12}:=\int_0^1 v_1\ol v_2dx$.
Below we always assume  $v\in \mH$.

\begin{theorem} \lb{T2}
Let $(\l,v')\in \C\ts \mH$ and $\t_j=\t_j(\l), j\in \N_3$ be the
multipliers of $\p(\l)$. Then we have
\[
\begin{aligned}
\lb{a1}   \t_3(\l)=e^{-i(\l-{\gb_3+o(1)\/2\l})},\qqq
\t_m(\l)=e^{i(\l-{\gb_m+o(1)\/2\l })}, \qqq m=1,2,
\end{aligned}
\]
\[
\begin{aligned}
\lb{a2}\tes
 \D_j(\l)=\cos (\l-{\gb_j+o(1)\/2\l }),  \qq j\in \N_3
\end{aligned}
\]
where $|\l-\pi n|\ge \d, $ for all $n\in \Z$ and for each $\d\in
(0,{1\/2})$, and
\[
\lb{a3}
\begin{aligned}
 \tes \gD(\l)={1\/(4\l)^2}(\gb_o+o(1))\sin^2(\l-k_{32})
\sin^2(\l-k_{31}),
\\
{\rm where} \qqq \tes  \qq k_{3m}={\gb_3+\gb_m+o(1)\/4\l}, \qq m=1,2
\end{aligned}
\]
 as $|\l|\to \iy$, uniformly on bounded subsets of $(\arg \l,v)\in [0,2\pi]\ts \mH$.

 \end{theorem}

Define a disc $\dD_r(z)=\{\l\in \C: |\l-z|<r\}, r>0,z\in \C$ and let
$\dD_r=\dD_r(0)$.

\begin{theorem} \lb{T3}
Let $\gb_o>0$ and $\|v'\|\le r<m$ for some $(v', r,m)\in \mH\ts\R_+\ts \N$. Then

\no i) $\gD (\l)$ has exactly $2+8m$ roots, counted with
multiplicities, in the disc $\dD_{\pi m+{\pi\/2}}$ and  exactly 4
real roots $\l_{n,j}^\pm, j=1,2$ in each disc $\dD_{\vs}(\pi n),
n>|m|$. There are no other roots.
\\
ii)  The function $\gD$ is positive on each interval $[\pi
(n-1)+\vs, \pi n-\vs]$ for $n$ large enough and $\vs>0$ small
enough.
\\
iii)  The branch points $\l_{n,j}^\pm\in \dD_{\vs}(\pi n), n>|m|,
j=1,2$ as the solutions of the equations $\t_3(\l)=\t_{j}(\l), \l\in
\dD_1(\pi n)$  have asymptotics uniformly on bounded subsets of $\mH$:
\[
\lb{ask3} \tes
 \l_{n,j}^\pm=\pi n+{\gb_3+\gb_j+o(1)\/4\pi n}\qqq \as \qq n\to \pm \iy.
\]
\end{theorem}

\no  Due to  Theorem \ref{T3}, we define the gaps
$\g_{n,j}=(\l_{n,j}^-,\l_{n,j}^+)$ in the spectrum $\gS_3$ by
\[\lb{sm3}... <\l_{1,1}^-\le \l_{1,1}^+<\l_{1,2}^-\le
\l_{1,2}^+<\l_2^-<...< \l_{n,1}^-\le \l_{n,1}^+<\l_{n,2}^-\le
\l_{n,2}^+<.....,
\]
where $\l_{n,j}^\pm \to \pm\iy $ as $n\to \pm\iy$. It is
important that $\l_{n,j}^\pm$ are branch points for Lyapunov functions
or 2-periodic eigenvalues and their labeling is given by Theorem \ref{T3}:

\no $\bu $ If $|n|>m$, then the gap $\g_{n,j}=(\l_{n,j}^-,\l_{n,j}^+)$, where $\l_{n,j}^\pm, j=1,2$  satisfy \er{sm3}.

\no $\bu $  If $|n|\le m$, then $\l_{n,j}^\pm, j=1,2$  satisfy \er{sm3}
and maybe  some  gaps are empty $\g_{n,j}=\es$.

\begin{figure}
\tiny
\unitlength 1mm 
\linethickness{0.4pt}
\ifx\plotpoint\undefined\newsavebox{\plotpoint}\fi 
\begin{picture}(116.85,105.5)(0,0)
\put(6.25,71.9){\line(0,-1){58.4}}
\put(4.65,22.9){\line(1,0){107.6}}
\put(5.05,42.5){\line(1,0){105.4}}
\put(4.65,62.5){\line(1,0){105.4}}
\put(111.05,38.9){\makebox(0,0)[cc]{$\l$}}
\put(10.25,68.3){\makebox(0,0)[cc]{$\D(\l)$}}
\put(2.85,39.5){\makebox(0,0)[cc]{$0$}}
\put(3.05,19.1){\makebox(0,0)[cc]{$-1$}}
\put(2.85,59.3){\makebox(0,0)[cc]{$1$}}
\qbezier(9.25,24.9)(20.65,20.2)(26.45,27.9)
\qbezier(20.85,42.5)(30.45,33.5)(26.45,27.9)
\qbezier(23.45,57.5)(15.15,49)(20.85,42.5)
\qbezier(38.25,59.9)(32.05,65.9)(23.45,57.5)
\qbezier(38.25,59.9)(42.85,56.1)(37.85,28.3)
\qbezier(37.85,28.3)(36.05,19)(31.05,26.9)
\qbezier(31.05,26.9)(29.65,28.3)(34.65,29.9)
\qbezier(34.65,29.9)(54.35,37.9)(57.65,43)
\qbezier(57.65,43)(71.35,62.4)(57.45,62.3)
\qbezier(57.65,43)(45.75,62.4)(57.45,62.3)
\qbezier(80.65,29.9)(60.95,37.9)(57.65,43)
\qbezier(84.25,26.7)(85.65,28.3)(80.65,29.9)
\qbezier(77.45,28.3)(79.25,19)(84.25,26.7)
\qbezier(77.05,60.1)(72.45,56.1)(77.45,28.3)
\qbezier(77.05,60.1)(82.85,65.9)(91.45,57.5)
\qbezier(91.45,57.5)(100.75,49)(94.05,42.5)
\qbezier(94.05,42.5)(84.45,33.5)(89.95,28.1)
\qbezier(107.05,24.9)(97.65,20.2)(89.95,28.1)
\put(5.05,87.5){\line(1,0){105.4}}
\put(109,83){\makebox(0,0)[cc]{$\l$}}
\put(8.25,83.5){\makebox(0,0)[cc]{$0$}}
\put(10.25,102.5){\makebox(0,0)[cc]{$\gD(\l)$}}
\put(20.65,90.25){\makebox(0,0)[cc]{$\l_{1,1}^-$}}
\put(27,90.25){\makebox(0,0)[cc]{$\l_{1,1}^+$}}
\put(34,90.25){\makebox(0,0)[cc]{$\l_{1,2}^-$}}
\put(39.5,90.25){\makebox(0,0)[cc]{$\l_{1,2}^+$}}
\put(53,90.25){\makebox(0,0)[cc]{$\l_{2,1}^-$}}
\put(97,90.25){\makebox(0,0)[cc]{$\l_{3,2}^+$}}
\put(90,90.25){\makebox(0,0)[cc]{$\l_{3,2}^-$}}
\put(84,90.25){\makebox(0,0)[cc]{$\l_{3,1}^+$}}
\put(76,90.25){\makebox(0,0)[cc]{$\l_{3,1}^-$}}
\put(64,90.25){\makebox(0,0)[cc]{$\l_{2,1}^+$}}
\qbezier(18.5,87.5)(20.75,84.5)(28,87.5)
\qbezier(28,87.5)(30.5,88.75)(31,87.5)
\qbezier(31,87.5)(36.875,83.5)(40.75,87.5)
\qbezier(40.75,87.5)(47.5,95.375)(51.25,87.5)
\qbezier(51.25,87.5)(55.625,76.375)(58,75.75)
\qbezier(64.75,87.5)(60.375,76.375)(58,75.75)
\qbezier(75.25,87.75)(68.5,95.375)(64.75,87.5)
\qbezier(84.5,87.5)(79.125,83.5)(75.25,87.5)
\qbezier(88,87.5)(85.5,88.75)(84.5,87.5)
\qbezier(97.5,87.5)(95.25,84.5)(88,87.5)
\qbezier(18.5,87.5)(13.625,96.125)(4.25,98.25)
\qbezier(97.5,87.5)(102.375,96.125)(111.75,98.25)
\thinlines \put(6.25,79){\line(0,1){26.5}}
\put(4.25,3.5){\line(1,0){112.6}}
\put(100,5.75){\makebox(0,0)[cc]{$\gS_3$}}
\thinlines
\multiput(18.297,86.797)(.002404,-.995192){85}{{\rule{.4pt}{.4pt}}}
\multiput(97.047,86.547)(-.002404,-.995192){85}{{\rule{.4pt}{.4pt}}}
\multiput(27.547,87.047)(0,-.992788){85}{{\rule{.4pt}{.4pt}}}
\multiput(87.797,86.797)(0,-.992788){85}{{\rule{.4pt}{.4pt}}}
\multiput(31.047,86.547)(0,-.990385){85}{{\rule{.4pt}{.4pt}}}
\multiput(84.297,86.297)(0,-.990385){85}{{\rule{.4pt}{.4pt}}}
\multiput(40.547,87.047)(0,-.992788){85}{{\rule{.4pt}{.4pt}}}
\multiput(74.797,86.797)(0,-.992788){85}{{\rule{.4pt}{.4pt}}}
\multiput(51.047,86.797)(.004808,-.995192){85}{{\rule{.4pt}{.4pt}}}
\multiput(64.297,86.547)(-.004808,-.995192){85}{{\rule{.4pt}{.4pt}}}
\linethickness{4pt} \put(18.65,3.9){\line(1,0){9.2}}
\put(97.05,3.9){\line(-1,0){9.2}} \put(31.25,3.9){\line(1,0){9.8}}
\put(84.45,3.9){\line(-1,0){9.8}} \put(51.45,3.9){\line(1,0){8.2}}
\put(64.25,3.9){\line(-1,0){8.2}}
\end{picture}
\lb{fig1} \caption{\footnotesize The function $\gD$, the Lyapunov
function $\D$ and the spectrum $\gS_3$}
\end{figure}
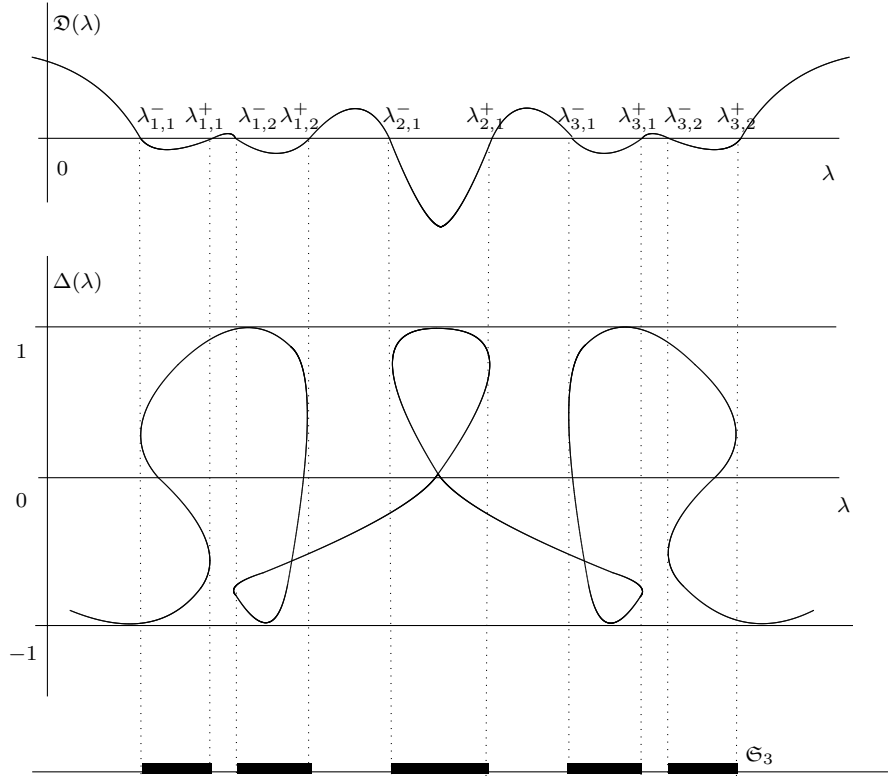


\begin{figure}
\tiny
\unitlength 0.8mm 
\linethickness{0.4pt}
\ifx\plotpoint\undefined\newsavebox{\plotpoint}\fi 
\begin{picture}(143,83.375)(0,0)
\put(3,70){\line(1,0){130}} \put(153,36){\line(0,1){.25}}
\qbezier[50](25.5,56.75)(40.5,83.375)(53.5,57.5)
\qbezier[50](25.5,56.75)(14.125,37.625)(6.75,32.5)
\qbezier[45](53.5,57.5)(61.75,40.875)(65,36)
\qbezier[55](65,36)(76.375,17.875)(84.25,18.75)
\qbezier[50](56.25,56.5)(44.875,37.375)(37.5,32.25)
\qbezier[50](56.25,56.5)(71.25,83.125)(84.25,57.25)
\qbezier[50](84.25,57.25)(92.5,40.625)(96.75,35.75)
\qbezier[45](70.25,30.5)(63.25,21.125)(54.25,20.25)
\qbezier[45](87,56.75)(82.75,49.25)(70.25,30.5)
\qbezier[50](87,56.75)(102,83.375)(115,57.25)
\qbezier[50](115,57.25)(123.25,40.875)(127.5,36)
\qbezier(23.5,57)(11.625,37.375)(4.25,32.25)
\qbezier(23.5,57)(38,83.125)(51,57.25)
\qbezier(51,57.25)(60.125,39.875)(59.5,31.25)
\qbezier(59.5,31.25)(59,25.625)(54.5,22.75)
\qbezier(58.175,57)(47.487,37.375)(40.85,32.25)
\qbezier(58.175,57)(71.225,83.125)(83.25,57)
\qbezier(83.25,57)(85,53)(78.75,43)
\qbezier(78.75,43)(74,35.625)(74.25,28.75)
\qbezier(74.25,28.75)(74.75,23.625)(81.25,20)
\qbezier(89.5,50.25)(90.75,46.625)(97,36.5)
\qbezier(89,56.5)(87.875,54.5)(89.5,50.25)
\qbezier(89,56.5)(104,83.375)(117,57.5)
\qbezier(117,57.5)(125.25,40.875)(129.5,36)
\put(15,36){\makebox(0,0)[cc]{$\D_1$}}
\put(44.75,30.5){\makebox(0,0)[cc]{$\D_2$}}
\put(58.25,17.75){\makebox(0,0)[cc]{$\D_3$}}
\put(38,74.75){\makebox(0,0)[cc]{$z_{1,n}$}}
\put(69.25,75.25){\makebox(0,0)[cc]{$z_{2,n}$}}
\put(100.25,75.75){\makebox(0,0)[cc]{$z_{3,n}$}}
\end{picture}
\lb{fig2} \caption{\footnotesize $\D_1$, $\D_2$, and $\D_3$ are
Lyapunov functions at high energy, $z_{1,n}$, $z_{2,n}$, and $z_{3,n}$ are periodic
eigenvalues. The dotted curve is the case where there are no
branching points . The solid curve is the case where there are 4
branching points}
\end{figure}
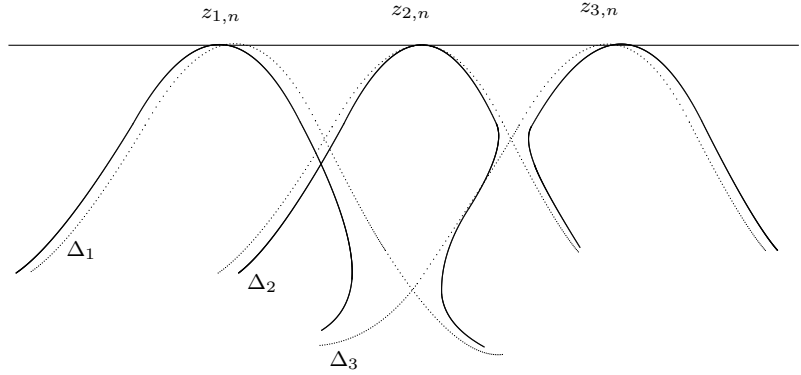

We discuss periodic and anti-periodic eigenvalues. Let
 $D_\pm(\l)=D(\pm 1,\l)$.   The zeros
of $D_{+}$ (or $D_{-}$) are the eigenvalues of  periodic (or
anti-periodic) problem for the equation $iJy'+Vy=\l y$. Let $\O=\Z\ts \N_3$.
Denote by $z_{2n,j}, (n,j)\in \O$ (and $z_{2n+1,j}$) the sequence of
zeros of $D_+$ (and $D_{-}$) counted with multiplicity such that $
z_{n,1}\le z_{n,2}\le z_{n,3}\le z_{n+1,1}\le z_{n+1,2}\le
z_{n+1,3}\le\dots . $ Note that $z_{n,j}$ is an eigenvalue of 2
periodic problem for the equation $iJy'+Vy=\l y$ and they have
asymptotics $z_{n,j}=\pi n+o(1)$ as $n\to \pm \iy$ and $j\in \N_3$.
If the spectrum $z_{n,j}, (n,j)\in \O$  is given,  then
we can recover $\gD$, the Riemann surface and the functions $T, \wt T$
see more in Theorem \ref{Tpe}.

\subsection{Conformal mappings and constants of motion}

A real function $f$ on the real line is upper semi-continuous if,
for any $x,r\in \R$ satisfying $f(x) < r$, there is a neighborhood
$U\ss\R$ of $x$ such that $f(y) < r$ for every $y \in U$. Let
$C_{us}$ denote the class of all real upper semi-continuous
functions  $h:\R\to [0,+\iy)$. With any $h\in C_{us}$  we associate
the "upper" domain
$$
\K(h)=\Big\{k=p+iq\in\C: q>h(p), p\in \R\Big\}\ss \C_+.
$$
Due to Theorems \ref{T2}, \ref{T3}  for each $j\in \N_3$ we
define the quasimomentum (some branch) by
\[
\lb{dkj}
\begin{aligned}
\ca  k_j(\l)=\arccos \D_j(\l),\ \ \ \ \Im \l\ge \l_c:=1+\sup_{n\in
\Z}|\Im \l_n^+|
 \\
k_j(\l)=\l+o(1) \qq \as \qq \Im \l\to +\iy \ac.
\end{aligned}
\]
 Define an {\bf averaged quasimomentum}
$\dk$ (shortly AQ), real function $\gp, \gq$ by
\[
\lb{dak} \dk(\l)=\gp(\l)+i\gq(\l)={1\/3}\sum_1^3 k_j(\l), \ \ \
\gq(\l)=\Im \dk(\l), \ \  \Im \l\ge \l_c.
\]
We present results about conformal mappings and identities, which will be important  to estimate spectral parameters for MS, similar to the KdV equation, see \cite{KK95}, \cite{K98}, \cite{K97}, \cite{K06}   and for the scalar NLSE \cite{K96}, \cite{K01}, \cite{K05}. We recall results about the
conformal mapping and trace formulas from \cite{K24}.
Define the domain
$
\mD_r=\{\l\in \C_+: \Im \l\ge r|\Re \l|\}, \qq r>0.
$

\begin{theorem}
\lb{Tk} Let $\dk={1\/3}\sum_1^{3} k_j$ be given by \er{dak}  for
some $v\in \mH$. Then
\\
i) The function $\dk$  has an analytic continuation from $\{\Im
\l\ge \l_c\}$ into $\C_+$, which is a conformal mapping from $\C_+$
onto $\dk(\C_+)=\K(h)\ss \C_+$ for some $h\in C_{us}$.
\\
ii) Let $\l=\x+i\n\in\C_+$. Then the following relations hold true:
 \[
 \lb{T23-1}
\dk(\l)=\l+{1\/\pi}\int_{\R}{\gq(t)\/t-\l}dt, \ \ \
\]
 \[
 \lb{T23-2}
\dk(\l)=\l-{Q_0+o(1)\/\l}
\]
as $ \l\in \mD_r$, $ |\l|\to \iy$,  for any $ r>0$, and
 \[
 \lb{T23-3}
Q_0={1\/\pi}\int_{\R}\gq(\x)d\x={1\/\pi}\iint_{\C_+}|\dk'(\l)-1|^2d\n
d\x+{1\/\pi}\int_{\R}\gq(\x)d\gp(\x)={2\/3}\|v\|^2,
\]
\[
 \lb{T23-4}
\tes
 \gq|_{\gS_3}=0,\qqq 0<\gq|_{\gS_1}\le {2\/3}\|v\|.
\]

\end{theorem}

We consider the conformal mapping for $v'\in \mH$ and the constants
of motion defined by
\[
\begin{aligned}
\lb{cm} \cH_0=\|v\|^2, \qq \cH_1=-i\lan v',v\ran, \qq \cH_2={1\/2}(\|v'\|^2+\|v\|_4^4),
\end{aligned}
\]
where $\lan \cdot,\cdot\ran$ is the scalar product in $\mH$ and $\cH_2$ is the Hamiltonian for the VNLS equation.

\begin{theorem} \lb{T5}
 Let $v'\in \mH$ and $\l=\x+i\n$. Then the AQ $\dk={1\/3}\sum_1^{3} k_j$ has asymptotics
\[
\lb{aqm1} \tes
\dk(\l)=\l -{Q_0\/\l} -{Q_1\/\l^2}-{Q_2+o(1)\/\l^3},
\]
as $\l\in \mD_r, |\l|\to \iy$  for any $ r>0$, where
$Q_{n}={1\/\pi}\int_{\R}\x^{n}\gq(\x)d\x, n=0,1,2$ and
 \[
\begin{aligned}
\lb{aqm2} & Q_1={2\/3}\cH_1,
\\
& Q_2={1\/\pi}\iint_{\C_+}|(\l(\dk(\l)-\l))'|^2d\n
d\x+{1\/\pi}\int_{\R}\x^2 \gq(\x)d\gp(\x)-{Q_0^2\/2}={2\/3}\cH_2.
\end{aligned}
\]

\end{theorem}

Asymptotics \er{aqm1} and identities \er{aqm2}   are similar to  \er{ask1}, \er{A0}-\er{A22} for ZS-systems.
{\it Our conjecture: the action variables for the periodic vector NLS equations are given by
$$
\A_\o ={2\/\pi}\int_{\g_\o}\gq(\l)d\l,\qq \o=(n,j)\in \O:=\Z\ts \N_3.
$$
There are estimates of  $\cH_2$ in terms of  $\sum n^2|\g_{n,j}|^2$ (and  $\sum n^2\A_{n,j}$)}. Note that $\|v\|^2=\sum \A_{n,j}$.

\begin{theorem} \lb{T6}
 Let $v'\in \mH$ and $\t_\bu=\max
\{1,  {\t_+\/2} \}$, where $\t_+:=\sup_{(\l,j)\in \O }|\t_j(\l)|$. Then
 \[
\begin{aligned}
\lb{H1} 
\tes
\cH_2\le {\|v\| \/3\pi}\sum_{\o\in \O,
\l_{\o}^\pm\in \R }|\g_{\o}|(|\l_{\o}^+|+|\l_{\o}^-|)^2,
\end{aligned}
\]
\[
\begin{aligned}
\lb{H2} 
\tes
\cH_2\le {\big(2 \t_\bu^4\big)^{1\/6}\/3\pi}\sum_{\o\in \O,
\l_{\o}^\pm\in \R }|\g_{\o}|^{7\/6}(|\l_{\o}^+|+|\l_{\o}^-|)^2.
\end{aligned}
\]
\end{theorem}

\no {\bf Short review.}
There is an enormous literature on  Hill operators and periodic ZS-operators,
including the inverse spectral theory, see  \cite{GT87}, \cite{IM75},
\cite{KK97}, \cite{K99},    \cite{K03}, \cite{K05}, \cite{MO75} and references therein. In order to get trace
formulas and estimates of potentials in terms of gap-lengths, spectral bands,
effective masses, action variables of Hamiltonians of the KdV, NLS equations   we study conformal
mappings constructed in \cite{Fi75}, \cite{MO75}, associated with the corresponding quasimomentum.
Various properties of such mappings, including trace formulas, were
obtained in \cite{KK95},  \cite{K97}, \cite{K05}, \cite{K06}, \cite{MO75}. The trace formulas are crucial to estimate  potentials (and the corresponding
Hamiltonians) in terms of gap-lengths, action variables. It was
obtained by Korotyaev in series of papers \cite{K96}, \cite{K98}, \cite{K05},
\cite{K06}.  Recall that Hilbert \cite{H09} obtained the first result about the conformal mappings from a multiply
connected domains onto a domain with parallel slits.

There are lot of results about periodic systems, see \cite{CG02}, \cite{C00}, \cite{GL55},    \cite{CK06q}, \cite{K08}, \cite{K10}, \cite{Kr55}, \cite{YS75} and  references therein. Such systems create new problems: Riemann
surfaces are multi-sheeted and their branch points can be real and complex;  spectral bands can have various multiplicity and their end points are
branch points or periodic eigenvalues. Note that the control of branch points is not simple. Until now the inverse problem for systems on the finite interval  is solved only for the Schr\"odinger operator with matrix-valued potential  \cite{CK06q}, \cite{CK09} (including characterization) and for first order systems
\cite{M99}, \cite{M15} (only uniqueness). The averaged quasi-momentum as a conformal mapping was constructed in \cite{CK06q}, \cite{K08}, \cite{K10}.

There are results about
the spectral theory for Manakov  systems, mainly devoted to inverse scattering problems associated with the MS on the real line \cite{AP04}.  
There are few papers about periodic Manakov  systems (see \cite{K10}). We mention a recent article \cite{K24}, where varuious results are obtained for $v\in\mH$ (it is described in Sect. 2).
In fact, our paper is a continuation
of \cite{K24} for the case $v'\in\mH$. For example, in this case we determine geometry
of the Riemann surface.

 Finally we discuss  3-order scalar periodic operators, see e.g.,
 \cite{BK24}, \cite{BK13}, \cite{BK12},  \cite{Mc81} and describe  their
properties. Such operators are used to integrate  the periodic
Boussinesq equation \cite{Mc81}. Here a Lax operator can be self-adjoint
or non self-adjoint. In both cases the Riemann surfaces are only 3-sheeted.
In the self-adjoint the spectrum the Lax operator
covers the real line and there are only a finite number of bands with multiplicity three \cite{BK12}, \cite{BK13}. Note that in order to study periodic good Boussinesq equations we need to consider non-self adjoint third order periodic
operators. Various properties of such operators (inverse problems,
divisors and so on) were discussed in \cite{Mc81}, \cite{BK24}.

\section {Spectral properties of $\cL$ for the case $v\in \mH$ \lb{Sec2}}
\setcounter{equation}{0}

\subsection{Properties of matrices }
We begin with some notational
convention. A vector $h=(h_n)_1^d\in \C^d$ has the Euclidean norm
$|h|^2=\sum_1^d|h_n|^2$, while a $d\ts d$ matrix $A$ has the
operator norm given by $|A|=\sup_{|h|=1} |Ah|$. Note that $|A|^2\le
\Tr A^*A$.
\\
$\bu$  We define the space $\M^{n,m}$ of all $n\ts m$ complex
matrices and let $\M^{d}=\M^{d,d}$.
\\
$\bu$ For any matrices $A, B\in \M^{d}$ the following identities
hold true:
\[
\lb{AB}
 \Tr AB=\Tr BA, \ \ \ \ \ \ \ \ \ol{\Tr A}=\Tr A^*.
\]

\no $\bu$ If $A\in \M^d$, then the following identity holds true:
\[
\lb{Das}\tes  \det (I+A)=\exp \Big(\Tr A-\Tr {A^2\/2}+\Tr
{A^3\/3}+..\Big)\ \ \as \qq \|A\|<1.
\]
$\bu$ Let $\mA$ be the class of all matrices $A=\ma
0_1&a_1\\a_2&0_2\am$, where $a_1,a_2^* \in \M^{1,2}$ and $0_d=0\in
\M^d$. Below we need following  identities for any  $A,B, C\in\mA$
and $\l\in \C$:
\[
\lb{23} J A=-AJ, \ \ \ \ \ \
e^{\l J} A=Ae^{-\l J},
\]
\[
\lb{213} ABC,\ JA ,\ e^{\l J}A, \ A^{2n-1} \in \mA,\qqq \forall \
n\in \N,
\]
\[
\lb{214} AB=\ma a_1b_2&0\\0&a_2b_1\am ,
\]
\[
\lb{215} \Tr A=0,\ \   \Tr JA^n=0,\ \  \forall \ n\in \N,
\]
\[
\lb{217}    e^{-i\l J}=\cos \l-iJ\sin \l.
\]
We need simple results.

\begin{lemma}  \lb{TvV}
i) The matrix $V_x$ has eigenvalues $z_\pm =\pm |v(x)|, z_o=0, $ and
\[
|V_x|=|v(x)|, \qq \forall \ x\in \T.
\]
ii) If $v'\in\mH$, then  the matrix  $W=iJV^2-V'$ has an estimate
\[
|W_x|\le |v(x)|^2+|v'(x)|, \qq \forall \ x\in \T.
\]
ii)  Let $\gc_j=\int_0^1|v_j|^2dx, j=1,2$. Then the matrix
$\mV=\int_0^1V_x^2dx$  has the form true:
\[
\lb{V22xx}
\begin{aligned}
\mV=\int_0^1V_x^2dx=\gc_o \os \gc,\qq
           \gc= \ma    \gc_1 & \ol \gc_{12} \\
           \gc_{12} & \gc_2 \am,
           \qq
           \Tr \gc= \|v\|^2=\gc_o,
\end{aligned}
\]
where $
\gc_{12}=\int_0^1 v_1\ol v_2dx$. Moreover, the matrix $\mV$ has
eigenvalues $\gb_1\le \gb_2\le \gb_3$ given by:
\[
\lb{V23} \tes \gb_3=\gc_o,\qq \gb_2={1\/2}(\gc_o+\sqrt{\gb_o}), \qquad
  \gb_1={1\/2}(\gc_o-\sqrt{\gb_o})   \ge
0,\qq \gb_o=(\gc_1-\gc_2)^2+4|\gc_{12}|^2.
\]

\end{lemma}

\subsection{Zakharov-Shabat operators } Consider a defocusing periodic NLS equation $iu_t=-u_{xx}+ 2|u|^2u$ with a scalar function $u\in L^2(\T)$. To integrate this
equation Zakharov and Shabat introduced the ZS-operator $\cL_{zs}$
on $L^2(\R,\C^2)$ given by
\[\lb{zs}
\cL_{zs}=iJ_{zs} {d\/ dx}+ V_{zs},\ \ \ \  J_{zs} =\ma 1&0\\ 0&-1\am, \
\ \ \ \ V_{zs}= \ma 0 &  \ol u\\  u & 0\am,\qq u\in L^2(\T).
\]
 Recall known facts about the ZS-operator (see, e.g.,
\cite{LS91}). The spectrum of $\cL_{zs}$ are purely absolutely
continuous and is a union of bands $(\l_{n-1}^+,\l_n^-)$ separated by the gaps $(\l_n^-,\l_n^+), n\in\Z$:
\[
\lb{zs12}
\s(\cL_{zs})=\R\sm \cup \g_n, \qq \g_n=(\l_n^-,\l_n^+),\qq
\l_n^\pm=\pi n+o(1)\ \ \as \ n\to \pm \iy.
\]
Here  $\l_n^\pm$ are eigenvalues of
 2-periodic problem for the equation $iJ_{zs}y_{zs}'+V_{zs}y_{zs}=\l y_{zs}$.
 The points   $\l_{2n}^\pm$ are eigenvalues of
 1-periodic problem for the equation $iJ_{zs}y_{zs}'+V_{zs}y_{zs}=\l y_{zs}$.
The points   $\l_{2n+1}^\pm$ are eigenvalues of
 anti-periodic problem for the equation $iJ_{zs}y_{zs}'+V_{zs}y_{zs}=\l y_{zs}$. If $\g_n=\es$, then $\l_{n}^\pm$ is a double  eigenvalue.
Introduce the $2\ts 2$-matrix valued solutions $y_{zs}(x,\l)$ of the
ZS-system
\[
\lb{zs1} iJ_{zs}y_{zs}'+V_{zs}y_{zs}=\l y_{zs},\qqq
y_{zs}(0,\l)=\1_{2}.
\]
The $2\ts 2$ matrix valued solutions $y_{zs}(1,\l)$ is entire and  has two
multipliers $\t_{zs}, \t_{zs}^{-1}$, which are analytic on the
2-sheeted Riemann surface. The Lyapunov function has the form $
\D_{zs}(\l)={1\/2}(\t_{zs}(\l)+\t_{zs}^{-1}(\l))$, is entire and
satisfies $\D_{zs}(\l)={1\/2}\Tr y_{zs}(1,\l) $. It defines the
band-gap structure of the spectrum.
 Moreover, we have
 $
 \D_{zs}(\l_{n}^\pm)=(-1)^n$ for all $n\in\Z.
 $
We introduce the Riemann surface $\cR_{zs}$ for the ZS-operator. For
the function $(1-\D_{zs}^2(\l))^{1\/2}$ $(\l\in \ol\C_+)$, we fix
the branch by the condition $(1-\D_{zs}^2(\l+i0))^{1\/2}>0$ for
$\l\in \s_1=[\l^+_{0},\l^-_1]$, and introduce the two-sheeted
Riemann surface $\cR_{zs}$ of $(1-\D_{zs}^2(\l))^{1\/2}$ obtained by
joining the upper and lower rims of two copies of the cut plane
$\C\sm\s_{ac}(\cL_{zs})$ in the usual (crosswise) way.
For each $u$ there exists a unique conformal mapping (the quasi-momentum)
$k:\C\sm\cup \bar \g_n\to \cK $ such that 
$$
\cos k(\l)= \D_{zs}(\l),\ \  \l\in \C\sm\cup \bar \g_n, \ \ \ \
\cK=\C\sm\cup (\pi n-ih_n,\pi n+ih_n),
$$
$$
k(\l)=\l-{\|u\|^2+o(1)\/ 2\l}\ \ \as \ \ \ \ \l\to i\iy , \ \ {\rm
and }\ \ \ \ k(\l)=\l+o(1)\qq \as \ \ \l\to \pm\iy,
$$
see \cite{Mi78}, \cite{KK95},
 where  the height $h_n\geq 0$ is defined by the equation $\ch h_n
=\max_{\l\in \g_n} |\D_{zs}(\l)| \geq 1$. First classical results
about such mappings  was obtained by Hilbert \cite{H09} for a finite
number of cuts (see also \cite{J58}). For Hill operators such
conformal mappings were constructed  simultaneously by Firsova
\cite{Fi75} and by Marchenko-Ostrovski \cite{MO75}. Various
properties of such conformal mappings, including trace formulas were
determined in \cite{KK95}, \cite{K96}, \cite{K97}, \cite{K00},
\cite{K05}, \cite{K06} for Hill and ZS-operators. For example, for
ZS-operators the estimate of gap lengths in terms of potentials was
obtained by EK \cite{K05}:
\[
\lb{KZS} \tes
 {1\/ \sqrt 2}g\leq \|u\|\leq 2g\big(1+g\big),\qqq {\rm where }\qq g=(\sum
 |\g_n|^2)^{1/2}.
\]
 Moreover, the estimates of Hamiltonians for NLS equations in terms
of action variables and gap-lengths were determined in \cite{K05}. The NLS equation
has the Hamiltonian $H_2$ and other two integrals of motion $H_0$
and $H_1$ given by
$$
H_o=\int_0^1|u|^2dx,\qq H_1=-i \int_0^1u'\ol udx,\qq
H_2={1\/2}\int_0^1(|u'|^2+|u|^4)dx,
$$
where $u=u_1+iu_2$.  If $u, u'\in \mH$, then
the quasimomentum $k(\cdot)$ has asymptotics
\[
\begin{aligned}
\lb{ask1}
k(\l)=\l-{Q_0\/\l}-{Q_1\/\l^2}-{Q_2+o(1)\/\l^3}\qq \as \ \Im \l\to \iy,
\end{aligned}
\]
see \cite{K05}, where the functionals $Q_j={1\/\pi}\int_\R \l^j\Im
k(\l+i0)d\l\ge 0,j=0,1,2$. 
Define the action variables $\ga_n={2\/\pi}\int_{\g_n}\Im k(\l+i0)d\l\ge 0, n\in \Z$, where $\Im k(\l+i0)|_{\g_n}> 0, n\in \Z$.
Recall the important identities from
\cite{K05}:
\[
\lb{A0} H_o=\sum_{n\in\Z} \ga_n=2Q_o= {1\/\pi}\iint_\C
|k'(\l)-1|^2d\x d\n , \qq \l=\x+i\n,\qqq{\rm if} \qqq u\in L^2(\T),
\]
\[
\lb{A1} H_1=\sum_{n\in\Z} (2\pi n)\ga_n=4Q_1,\qqq \qqq\qqq \qqq{\rm
if} \qqq u,u'\in  L^2(\T),
\]
\[
\lb{A22} H_2=\sum_{n\in\Z} (2\pi n)^2\ga_n+2H_0^2-H_{2,1}=4Q_2,\qq
\qqq \qqq {\rm if} \qqq u,u'\in L^2(\T),
\]
where $H_{2,1}(u)$ is some function of $u\in L^2(\T)$. In fact, we
rewrite $H_o, H_1$ and the main part of $H_2$ in terms of simple
functions of the actions $\ga=(\ga_n)_{n\in \Z}$. Moreover,
$H_{2,1}(u)$ is well defined for $u\in L^2(\T)$ and the following
estimates  hold true (see \cite{K05}):
\[
\lb{eV} 0\le H_{2,1}(u)\le {4\/3}\|u\|^2,\qqq \qqq \forall  \ u\in
L^2(\T).
\]

\subsection{Fundamental solutions of Manakov systems}
We begin to study the fundamental $3\ts 3$-matrix solutions
$y(x,\l)$ of the equation
\[
\lb{DE1} iJy'+Vy=\l y,\qqq  y(0,\l)=\1_{3}\qq \forall \ \l\in\C.
\]
This solution $y$ satisfies the standard integral equation
\[
\lb{26} y(x,\l)=y^o(x,\l)+i\int_0^x e^{i\l (s-x)J}JV_sy
(s,\l)ds,\ \ \ y^o(x,\l)=e^{-i\l xJ}.
\]
It is clear that equation  \er{26}  has a solution as a power series
in $V$ given by
\[
\lb{27}
\begin{aligned}
&  y=y^o+\sum_{n\ge1}y_n, \ \ \
y_1(x,\l)=i\int_0^xe^{i\l (s-x)J}JV_se^{-i\l sJ}ds=
i\int_0^xe^{i\l (2s-x)J}JV_sds, 
 \\
&  y_n(x,\l)=i\int_0^xe^{i\l (s-x)J}JV_sy_{n-1}(s,\l)ds,\ \ n\ge2,
\end{aligned}
\]
where we have used \er{27}, \er{23}.  Recall  the basic results about the function $y$.

\no  \begin{lemma}  \lb{Tfs} For each $(\l,v)\in \C\ts \mH$ there
exists a unique solution $y(x,\l)=y(x,\l,v)$ of the equation
\er{26} given by \er{27} and series \er{27} converge uniformly on
bounded subsets of $\R\ts\C\ts \mH$. For each $x\in [0,1]$ the
function $y(x,\l,v)$ is entire on $\C\ts \mH$ and  for any
$(n,x,\l)\in \N\ts [0,1]\ts \C$ satisfies:
\[
\lb{217x}
\begin{aligned}
 |y_n(x,\l)|\le e^{x|\Im \l|}{\|v\|^n\/n!},
 \qqq
 |y(x,\l)|\le e^{x|\Im \l|+\|v\|},
 \end{aligned}
\]
where $\|v\|^2=\int_0^1(|v_1(x)|^2+|v_2(x)|^2)dx$ and
\[
\lb{218} |y(x,\l)-\sum_0^{n-1}y_j(x,\l)|\le {\|v\|^n\/n!}e^{x|\Im
\l|+\|v\|},
\]
\[
\lb{219} y(x,\l)-e^{-i\l xJ}=o(1)e^{x|\Im \l|}\ \qqq as \qq |\l|\to
\iy,
\]
uniformly on bounded subsets of $\R\ts \mH$. Moreover, if the
sequence $v^{(m)}\to v$ weakly in $\mH$, as $m\to \iy$, then $y
(x,\l,v^{(m)})\to y(x,\l,v)$ uniformly on bounded subsets of
$[0,1]\ts \C$.
\\
Moreover,  for any fixed $\d>2$ the following asymptotics
\[
\lb{aspzn}
\begin{aligned} y(x,\l_n)=y_o(x,\l_n)+y_1(x,\l_n)+y_{(2)}(x,\l_n),
\\
|y_1(x,\l_n)|=\ell^{\d}(n) ,\qq |y_{(2)}(x,\l_n)|=\ell^{\d/2}(n),
\end{aligned}
\]
as $n\to \iy $ hold, uniformly on $[0,1]\times\{|\l_n-\pi n|\leq
{1\/4}\}\ts \{\|v\|\le r\}$ for any $r>0$.

\end{lemma}

We describe the functions $T(\l)=\Tr \p(\l)$, where
$\p(\l)=y(1,\l)$. From \er{27} we have $\p
(\l)=\sum_{n\ge0}\p_n(\l)$ and then $T(\l)=\sum_{n\ge0}T_{2n}(\l)$,
where $T_n:=\Tr \p_n$ and $T_{2n+1}=0$.

\begin{lemma}
\lb{Tmo} The function $T(\l)=\Tr \p(\l)$ is entire in $\l\in \C$
and satisfies
\[
\lb{223} T(\l, v)=T(\l,-v)=\sum _{n\ge 0}T_{2n}(\l,v),
\]
\[
\lb{220} |T_{2n}(\l,v)|\le 2e^{|\Im \l|}{\|v\|^{2n}\/(2n)!},
\qqq
 |T(\l,v)|\le 4e^{|\Im \l|}\ch \|v\|,
\]
\[
\lb{226} T(\l,v)=T_0(\l)+o(e^{|\Im \l|})\ \  \as \ \
|\l|\to \iy.
\]
Series \er{223} converge uniformly on bounded subsets of $\C\ts
\mH$. If a sequence $v^{(m)}$ converges weakly to $v$ in $\mH$ as
$m\to \iy$, then $T(\l,v^{(m)})\to T(\l,v)$ uniformly on bounded
subsets of $\C$.
\end{lemma}

The multipliers  $\t_1,\t_2, \t_3$,  trace $T$ and $\wt T(\l)=\ol T(\ol \l)$ have
the following properties (see \cite{K24})
\[
\lb{4}
\begin{aligned}
&  \det \p(\cdot)=e^{i\l}=\t_1\t_2\t_3,
  \\
& \tes  \wt
T={1\/\t_1}+{1\/\t_2}+{1\/\t_3}=e^{-i\l}(\t_1\t_2+\t_2\t_3+\t_3\t_1).
\end{aligned}
\]
We recall the properties of multipliers and Lyapunov functions (see
p.109 \cite{YS75}, \cite{K10}).

\begin{theorem} \lb{TL}
{\bf (Lyapunov Theorem).} Let $v\in \mH$. Then
\\
\no i) If $\t=\t(\l)$ is a multiplier of $\p(\l)$  for some $\l\in
\C$, then $1/\ol \t(\l)$ is a multiplier of $\p(\ol \l)$. In
particular, if $\l\in \R$, then ${1/\ol\t(\l)}$ is the multiplier of
$\p(\l)$.
\\
ii) If $\t(\l)$ is a simple multiplier and $|\t(\l)|=1$ for some
$\l\in\R$, then $\t'(\l)\ne 0$.
\\
iii) Let some $\D_j, j\in \N_3$ be real analytic on some interval
$I=(a,b)\ss\R$
 and $\D_j(I)\ss (-1,1)$. Then $\D_j'(\l)\ne 0$ for each $\l\in I$
(the monotonicity property).

\no v) Each gap $(\l^-,\l^+)$ in the spectrum $\gS_3$ is the
interval $\ss \gS_1$ and its edges are periodic (anti-periodic)
eigenvalues or branch points of $\D$.

\end{theorem}

 For a matrix-valued (and $\C$-valued) function $f(x,\l)$ we formally define the mapping by
$$
f(x,\l)\to \wt f(x,\l)=f^*(x,\ol \l),\qq (x,\l)\in \R\ts \C,
$$
and   the Wronskian for the functions $f,g$:
\[
\{f,g\}=\wt fJg.
\]
If $f, g$ are solutions of the equation $iJ\p'+V\p=\l\p$, then the
Wronskian $\{f,g\}$ does not depends on $x$. In particular, the
Wronskian $\{y,y\}=J$ is a constant matrix and we have
\[ \lb{3}
\{y,y\}=\wt \p J\p=J \qq \Rightarrow \qq \p^{-1}=J\wt \p J.
\]

\no Recall results about the discriminant
$\gD=-{e^{-i2\l}\/4^3}(\t_1-\t_2)^2(\t_2-\t_3)^2(\t_1-\t_3)^2 $ from
\cite{K24}. It has having the form
\[
\lb{rzz} \gD=-{1\/64}\Big(T^2\wt T^2-4e^{-i\l }T^3-4e^{i\l } \wt
T^3+18T\wt T-27\Big).
\]

\begin{lemma}  \lb{Tdd} i) Let
multipliers $\t_j=e^{i\f_j}, j\in \N_3$ for some $ \f_j\in\C$. Then
the function $\cD(\l)$ satisfies
\[
\lb{rqm}
\begin{aligned}
\tes \gD=\sin^2{\f_1-\f_2\/2}\sin^2{\f_1-\f_3\/2}\sin^2{\f_2-\f_3\/2}.
\end{aligned}
\]
\[
\lb{di1} |\cD(\l)|\le 2^5e^{4|\n|+4\|v\|},\qq  \l\in \C.
\]
ii) Let the quasimomentum $\f_1(\l)=p_1(\l)+iq(\l)\in \C_+,\
\f_3(\l)\in \R$ for all $\l\in \g=(\l^-,\l^+)$, where
 $\g\ss \gS_1$ is some gap. Then on $\g$ we have
\[
\lb{rqmg}
\begin{aligned}
\tes -{\gD(\l)}={1\/4}(\sh^2{q})\Big(\ch q-\cos (p_1-\f_3)\Big)^2\le
\sh^2{q}  \ch^2{q} \le {e^{4q}\/2^4}\le {\t_+^4\/2^4}\le
{e^{4\|v\|}\/64},
\end{aligned}
\]
\[
\lb{rqmb}
\begin{aligned}
 \tes {q^6\/16}<\ {1\/4}(\sh^2{q})\Big(\ch q-1)\Big)^2\le -{\gD(\l)}\le
 {|\g|\/8}\t_\bu^4,
\end{aligned}
\]
where $\t_+:=\sup_{(\l,j)\in \R\ts \N_3 }|\t_j(\l)|$ and $\t_\bu=\max
\{1,  {\t_+\/2} \}$. Moreover, we have
\[
\lb{Dm} \tes \sup_{\l\in \R} |\gD(\l)|\le \t_\bu^4.
\]

\end{lemma}

Define a set $E_+$ of all periodic $z_{2n,j}$ and a set $E_-$ of all
anti-periodic eigenvalues $z_{2n+1,j}$ by
$$
E_+=\{z=z_{2n,j}, (n,j)\in \O\},\qqq E_-=\{z=z_{2n+1,j},
(n,j)\in \O\}, \qq \O=\Z\ts \N_3.
$$
Let $z_{n,j}^o=\pi n, (n,j)\in \O$ be 2-periodic eigenvalues
for $v=0$. Introduce the spaces $\ell^p, p\geq 1$ of sequences
$f=(f_n)_{n\in\Z}$ equipped with the norm $\|f\|_{p}^p=\sum
|f_n|^p$. Below we write $a_n=b_n+\ell^\d(n)$ iff the sequence
$(a_n-b_n)_{n\in \Z}\in \ell^\d$ for some $\d\ge 1$.

\begin{theorem}   \lb{Tpe}
The eigenvalues  $z_{n,j}, (n,j)\in \O$  have asymptotics as
$n\to\pm\iy$:
 \[
 \lb{a8}
 \begin{aligned}
& z_{n,j}=\pi n+\z_{n,j}+\ell^\d(n),\qqq  {\rm where }\qq
\z_{n,2}=0, \qq \z_{n,1}=|\wh v(\pi n)|=-\z_{n,3},
\end{aligned}
\]
and   $\d>1$ and $ \wh v(\l)=\int_0^1 e^{i2\l x}v(x)dx$, uniformly on bounded subsets of $\mH$.
Moreover, let $D_\pm=D(\pm 1,\cdot)$ and $D_\pm(0)\ne 0$. Then the
functions $D_\pm$  have the Hadamard factorizations
\[
\lb{Dpm1}
\begin{aligned}
\tes D_\pm(\l)=D_\pm(0)e^{i{\l\/2}} \lim_{r\to +\iy}\prod_{|k_n|\leq r,  k_n\in
E_\pm}\Big(1-{\l\/k_n}\Big),
\end{aligned}
\]
where the constants  $D_+(0)=i2 \Im T(0),\  D_-(0)=2(1+ \Re T(0))$
and the product converges uniformly in every bounded disc. If the
spectrum $E_\pm$ and the constants $D_\pm(0)$ are given,  then
we can recover $\gD$, the Riemann surface and the functions $T, \wt T$ by
\[
\lb{fjd1}
\begin{aligned}
\tes T={1\/2}(D_-+D_+)-e^{i\l},\qqq
  \wt T=e^{-i\l}{1\/2}(D_--D_+)-e^{-i\l}.
\end{aligned}
\]

\end{theorem}

\subsection{The Lyapunov function}

In order to study the Lyapunov function we  introduce the matrix-valued function
$\L={1\/2}(\p+\p^{-1})$ with eigenvalues
$\D_j={1\/2}(\t_j+\t_j^{-1}),j\in \N_3$. Due to \er{3} we obtain
\[
\lb{} \L(\l)={1\/2}(\p(\l)+\p^{-1}(\l))={1\/2}(\p(\l)+J\wt\p(\l)J).
\]
Define the determinant $D_\L(\l,\a):=\det (\L(\l)-\a \1_{3}),
\l,\a\in \C$, which has the form
\[
\lb{f1}
\begin{aligned}
&   D_\L(\l,\a)=-\a^3+\a^2\cT(\l)-\a\cT_1(\l)+\det \L(\l),
\end{aligned}
\]
where
\[
\lb{e2} \tes  \cT=\Tr \L=\D_1+\D_2+\D_3={1\/2}(T+ \wt T),
\]
\[
\lb{f3} \tes   \cT_1={1\/4}(e^{i\l }T+1)(e^{-i\l }\wt T+1)-1,
\]
\[
\lb{f4} \tes   \det \L(\l)=\D_1\D_2\D_3={1\/8}(2\cos \l +e^{i\l }\Tr
\wt\p^2+ e^{-i\l }\Tr \p^2),
\]
\[
\lb{f5} \Tr \p^2=T^2-2e^{i\l }\wt T,\qqq \Tr \wt \p^2=(\wt
T)^2-2e^{-i\l }T.
\]
Define the discriminant $\gD_\D$ of the polynomial $D_\L(\l,\a)$ by
$$
\gD_\D=(\D_1-\D_2)^2(\D_1-\D_3)^2(\D_2-\D_3)^2.
$$
It has the form
\[
\lb{f6}
\begin{aligned}
 \gD_\D=\cT^2\cT_1^2-4D_o\cT^3-4\cT_1^3-18D_o\cT\cT_1-27D_o^2,
\end{aligned}
\]
where $D_o=\det \L$. We discuss the properties of $\gf, \gD_\D,
\gD$.

\begin{lemma}
\lb{TL1} Let $\gf=\cT_--\sin \l$, where  $\cT_-={1\/2i}(T-\wt T)$. Then
$\gf, \gD_\D$ are entire, real on the real line and satisfy
\[
\lb{dd1}
\begin{aligned}
D(e^{i\l},\l)=2ie^{i2\l}\gf(\l),\qq \forall \ \l\in \C,
\end{aligned}
\]
\[
\lb{dd2}
\begin{aligned}
\tes  4\gD \ \gf^2=\gD_\D.
\end{aligned}
\]
If the Riemann surface is 3-sheeted, then  each zero of $\gf$
belongs to $\gS_3$.
\end{lemma}

\subsection{Two sheeted Riemann surfaces}

Recall the set
$
\mH_o=\big\{v\in \mH: v=ue,\qq (u,e)\in L^2(\T)\ts \C^2, \qq
|e|=1\big\}.
$
 Let $v=ue\in \mH_o$ for some constant vector $e\in \C^2, |e|=1$  and $u\in L^2(\T)$.
 We show that $\cL$ is unitary equivalent to the operator $\cL_u=\cL_{zs}\os {d\/idx}$,
where $\cL_{zs}$ is the ZS-operator on $L^2(\R,\C^2)$ given by
\er{ddzs}. Let $\cU=1 \os U \in \M^3 $, where the matrix $U\in \M^2$
is  unitary and given by
 $$
 U^*=( e, e_o),\qq e_o\in \C^2,\qq |e_o|=1, \qq e_o \perp e.
 $$
 Define the operator $\cL_u= \cU \cL \cU^*$ and the vector $w=U v=uU e=(u,0)^\top$. We have
 \[
\lb{uzs1x}
\begin{aligned}
\cL_u= \cU \cL \cU^*=i{d\/dx}J +V_u ,\qqq J= \cU J \cU^*=J,\qqq V_u= \cU
V \cU^*.
\end{aligned}
\]
We compute the operator $V_u$:
$$
 V_u= \cU V \cU^*=\ma 1 & 0\\ 0 & U\am  \ma 0 & v^*\\ v &0\am   \ma 1 & 0\\ 0 & U^*\am
 =\ma 0 & v^*\\ Uv & 0      \am        \ma 1 & 0\\ 0 & U\am =
 \ma 0 & w^*\\ w & 0 \am =V_{zs}\os 0.
$$
Thus we have $\cL_u=\cL_{zs}\os {d\/idx}$ and $V_u=V_{zs}\os 0$. The
operators $\cL, \cL_u$ are unitary equivalent and their monodromy
matrices are unitary equivalent also. Then their multipliers,
periodic eigenvalues are the same. The multipliers and the Lyapunov
functions have the form
\[
\lb{mtt} \t_1=e^{i\l},\ \  \t_2=e^{ik}, \ \ \t_3=e^{-ik}, \ \
\D_1=\cos \l, \ \ \D_2=\D_3=\D_{zs}=\cos k(\l),
\]
 where $k=k(\l)$ is
the quasimomentum for the operator $\cL_{zs}$. Recall  that for
ZS-operators  the Lyapunov function $\D_{zc}=\cos k(\l)$.  We
compute $T$ and $\gD$:
\[
T=e^{-ik}+e^{ik}+e^{i\l}=2\cos k+e^{i\l}=2\D_{zc}+e^{i\l}
\]
and
\[
\lb{r0}
\begin{aligned}
& \tes
64\gD=-e^{-i2\l}(e^{-ik}-e^{ik})^2(e^{ik}-e^{i\l})^2(e^{i\l}-e^{-ik})^2
= 64 \sin^2{k} \sin^2{k-\l\/2}\sin^2{k+\l\/2}
\\
&=16\sin^2{k}(\cos k-\cos \l)^2=16(1-\D_{zs}^2(\l))(\D_{zs}(\l)-\cos
\l )^2.
\end{aligned}
\]
 Then from \er{sro} we have  that $
\gS_3=\s(\cL_{zs})=\{\l\in \R: \D_{zs}^2(\l)<1\}. $ We also have
zeros of the factor $(\D_{zs}(\l)-\cos \l )^2\ge 0$ on $\R$ and it
does not create the gaps in the spectrum $\gS_3$. Similar arguments
give
$$
\gf=0,\qq \gD_\D=0,\qq D_\pm=2(1\mp\D_{zs})(e^{i\l}-1).
$$
 Then from the identities
$D(\t,\l)=-(\t-\t_1)(\t-\t_2)(\t-\t_3)$ and \er{mtt}  we have
\[
D(\t,\l)=-(\t-e^{i\l})(\t^2-2\t \D_{zs}+1).
\]
 The Lyapunov function $\D_1=\cos \l$, which implies $z_{n,1}=\pi n, n\in \Z$.
The Lyapunov function $\D_2=\D_3=\D_{zs}$ and it gives the spectral
bands of $\gS_3$ and the corresponding gaps $\g_n=(\l_n^-,\l_n^+),
n\in \Z$. Here $\l_n^\pm$ are 2-periodic eigenvalues and they have
asymptotics
\[
\lb{RS23}
\begin{aligned}
 z_{n,2}=\pi n, \qq  z_{n,2}=\l_n^-,\qq z_{n,3}=\l_n^+,
 \qq
 \l_n^\pm =\pi n\pm |\wh v(\pi n)|+\ell^\d(n)
\end{aligned}
\]
as $ n\to \pm  \iy$, for some $\d>1$ and if $g=(\sum
|\g_n|^2)^{1/2}$, then we have estimates \er{KZS}.

 \begin{theorem}
\lb{T2sR} i)  Let an interval $\o\ss \g_o$
for some open gap $\g_o=(\l^-, \l^+)$ in the spectrum
$\gS_3$ for some $v\in\mH$.
 The following relations about the Riemann surface $\cR$ hold true:
\[
\lb{uzs2}
\begin{aligned}
v\in \mH_o \qq  \Leftrightarrow
\qq \cR\  {\rm is\  2-sheeted}\qq \Leftrightarrow \qq  \gD_\D=0 \qq  \Leftrightarrow\qq \gf=0
\\
\Leftrightarrow\qq \D_2(\o)\ss \R \sm[-1,1] \qq \Leftrightarrow \qq  \D_2=\D_3 \qq \Leftrightarrow \qq
\t_2\t_3=1 \qq \Leftrightarrow \qq  \t_1=e^{i\l}.
\end{aligned}
\]
or
\[
\lb{uzs3}
\begin{aligned}
 v\in \mH_o \qq  \Leftrightarrow
 \qq \cR\  {\rm is\  2-sheeted}\qq \Leftrightarrow \qq  \gD_\D=0 \qq  \Leftrightarrow\qq \gf=0
\\
\qq \D_1(\o)\ss \R \sm[-1,1] \qq \Leftrightarrow \qq  \D_1=\D_3 \qq \Leftrightarrow \qq
\t_1\t_3=1 \qq \Leftrightarrow \   \t_2=e^{i\l}.
\end{aligned}
\]
ii) Let $v\in \mH_o$, i.e., $v=ue$ for some constant vector $e\in \C^2,
|e|=1$  and $u\in L^2(\T)$. Then $\cL$ is unitary equivalent to the operator
$\cL_u=\cL_{zs}\os {d\/idx}$, where $\cL_{zs}$ is the Zakharov-Shabat operator
on $L^2(\R,\C^2)$ given by \er{ddzs}, and they satisfy \er{uzs1x}.
Furthermore, the monodromy operators for $\cL_u$ and $\cL$ are
unitary equivalent, their multipliers, periodic eigenvalues are the
same, the Riemann surface $\cR$ is 2-sheeted and
$\gS_3(\cL)=\s(\cL_{zs})$.

\end{theorem}
\no {\bf Remark.} 1) The two-sheeted Riemann surface of $\cL$ is the Riemann surface
for ZS-operators. Thus all their branch points are the 2-periodic eigenvalues
for ZS-operators.

\no 2) If the Riemann surface is two-sheeted, then we can solve the
inverse problem in terms of gap lengths, using the results of the
author \cite{K05} fo ZS-operators. Or via the Greber-Kappeler
results \cite{GK14} we can discuss action-angle bijection.

\section { The monodromy matrix and its trace for  $v'\in \mH$ \lb{Sec3}}
\setcounter{equation}{0}

\subsection{Monodromy matrices  for $v'\in \mH$}
In order to determine the asymptotics of the monodromy matrix $\p$
and other functions for $v'\in \mH$ we need some modification.
Define the integral operator $K$ and the matrix-valued function
$a_x(\l), x\ge 0$ by
\[
\lb{31} \tes (Kf)(x)=\int_0^x\!e_{s-x}W_sf(s)ds,\ \ \ W=iJV^2-V',\qq
\ a_x(\l)=(\1-\ve V_x)^{-1},
\]
where $\ve={1\/2\l}$ and $e_x=e^{ix\l J}$.
We consider the modified function $X=a^{-1}y a_0$ , which is a solution for
the integral equation with $y^o=e^{-ix\l J}$ given by
\[
\lb{33} \tes X=y^o+\ve Ka X,\qq \l\in \mB_v=\{\l:|\l|\ge \|v\|_{\iy}\}.
\]
\begin{lemma}  \lb{TaP}
Let $(\l,v')\in \mB_v \ts\mH$. Then the matrix-valued function
$X=a^{-1}y a_0$ is a solution for the integral equation given \er{33}
and for $\l\in \mB_v$ it has the form
\[
\lb{34} X=\sum_{n\ge 0}\ve^n X_n,\qqq  X_n= (Ka)^ny^o,\qq y^o=e^{-ix\l J}.
\]
 Moreover,  for any $n, j-1\in \N, \l\in \mB_v$
the following estimates hold true:
\[
\lb{35} |X_n(x,\l)|\le {2^ne^{|\Im \l|x}\/n!}\lt(\int_0^x
|W_s|ds\rt)^n,
\]
\[
\lb{36}
 |X(1,\l)-\p_0(\l)-\sum_1^{j-1}\ve^nX_n(1,\l)|\le
{\vk^j \/j!}e^{(|\Im \l|+\vk)},\qqq \vk:={\|v\|^2+\|v'\| \/|\l|}.
\]
Here  \er{34}-\er{36} hold true uniformly on bounded subsets of
$[0,\iy)\ts\mB_v\ts \mH$.
\end{lemma}

 \no {\bf Proof.} Let $y_x=y(x,\l)$ and $y_x^o=e^{-i\l xJ}$ for shortness.
Using \er{26},  $\ve={1\/2\l}$ and the identity
$$
iJ \big(e^{i\l xJ}y_x\big)'=e^{i\l xJ}(-\l y_x+iJy_x')=-e^{i\l
xJ}V_xy_x,
$$
and integrating by parts  we get
$$
\begin{aligned}
y_x-y_x^o=i\int_0^x Je^{i\l J(s-x)}V_sy_sds=
 \int_0^x \rt(\ve e^{i\l J(2s-x)}\rt)'_s
\rt(V_se^{i\l sJ}y_s\rt)ds
\end{aligned}
$$
$$
=\ve e^{i\l J(2s-x)}\rt(V_se^{i\l sJ}y_s\rt)\rt|_0^x- \ve\int_0^x
e^{i\l J(2s-x)}\rt(V_se^{i\l sJ}y_s\rt)'ds
$$
$$
=\ve \rt(V_xy_x-e^{-i\l xJ}V_0\rt)-\ve\int_0^t e^{i\l
J(2s-x)}\rt(V'_se^{i\l sJ}-ie^{-i\l sJ}JV_s^2\rt)y_sds.
$$
Thus we obtain
$$
(\1-\ve V_x)y_x=e^{-i\l xJ}(\1-\ve V_0)+\ve\int_0^x e^{i\l
J(s-x)}W_sy_sds, \qqq W=iJV^2-V',
$$
which yields \er{33} for $X=a_x^{-1}y_xa_0$ and $X_n= (Ka)^ny^o$. We will show
\er{34}-\er{36}. Using $|a_x(\l)|\le 2$ for $\|v\|_\iy\le |\l|$ and
$|y_x^o|\le e^{|\Im \l|x}$, we have
$$
|X_n(x,\l)|\le 2 \int_0^x e^{|\Im
\l|(x-x_1)}|W_{x_1}||X_{n-1}(x_1)|dx_1
$$
$$
\le
2^n\!\!\int_0^x\!\!dx_1\!\!\int_0^{x_1}\!\!\!\!dx_2\dots\!\!\!
\int _0^{x_{n-1}}\!\! e^{|\Im \l|x}|W_{x_1}|\dots |W_{x_n}| dx_1=
{2^n\/n!}e^{|\Im \l|x}\lt(\int_0^x |W_{t}|dt \rt)^n,
$$
which gives \er{34} and \er{35}. Estimates \er{35} and
$\int_0^1|W_x|dx\le \|V'\|+\|V\|^2$ imply  \er{36}. \BBox

Now we compute coefficients $\P_n:=X_n(1,\l)$  in the decomposition
\er{34} given by
$$
\P=X(1,\l)=\sum_{n\ge 0}\ve^n \P_n, \qqq \P_n:=X_n(1,\l).
$$
\no  1) Consider the first simple component $\P_1$. Below we need identities
\[\lb{ia}
a(\l)=(\1-\ve V)^{-1}=\1+\ve V+ \ve^2 V^2+..., \qqq \l\in \mB_v,
\]
\[
\lb{ia2}
e_{t}(iJV_t^2-V_t')e_{-t}=iJV_t^2-e_{2t}V_t',\qq
e_{t}(iJV_t^2-V_t')V_te_{-t}=ie_{2t}JV_t^3-V_t'V_t,
\]
where $e_t=e^{i\l tJ}$.  Thus using \er{ia} we get
$$
\begin{aligned}
X_1=Kay^o=(K+\ve KV+\ve^2 KV^2)y^o+O(\ve^3e^{
|\n|})
=e_{-1}(Y_1+\ve Y_{1,2}+\ve^2 Y_{1,3})+O(\ve^3e^{ |\n|}),
\end{aligned}
$$
where due to the identity \er{31} and the definition   $ \wh
{V_\l^3}:=\int_0^1\! e^{i2x\l J}V_x^3dx$,  we have
\[
\begin{aligned}
\lb{Y1} & Y_{1}=e_{1}Ky^o=\int_0^1(iJV_t^2-e_{2t}V_t')dt=iJ\mV-\wh
{V_\l'},\qqq \wh {V_\l'}:=\int_0^1\!\!\! e^{i2x\l J}V_x'dx,
\\
& Y_{1,2}=e_{1} KVy^o=\int_0^1e_{t}(iJV_t^2-V_t')V_te_{-t}dt=iJ\wh
{V_\l^3}-\int_0^1V_t'V_tdt,
\\
&
Y_{1,3}=e_{1}KV^2y^o=\int_0^1e_{t}(iJV_t^2-V_t')V_t^2e_{-t}dt=iJ\int_0^1
V^4_tdt-\int_0^1e^{i2t\l J}V_t'V_t^2dt.
\end{aligned}
\]
\no 2) Consider the second term $X_2=(Ka)^2\p_0$. Using \er{ia} we obtain
$$
X_2= K^2y^o+ \ve (KVKy^o+KKVy_0)+ O(\ve^2e^{|\n|})=e_{-1}
(Y_{2,1}+\ve (Y_{2,a}+Y_{2,b}))+ O(\ve^2e^{|\n|}).
$$
Consider the component $Y_{2,1}=e_{1}K^2y^o$. The identity \er{ia2} implies
\[
\begin{aligned}
\lb{Y2} & Y_{2,1}=e_{1}K^2y^o=
\int_0^1\!\!dt\!\!\int_0^t(iJV_t^2-e_{2t}V'_t)
(iJV_s^2-e_{2s}V_s')ds
\\
& =\int_0^1dt\int_0^t\rt[-V_t^2V_s^2-ie_{2s}JV_t^2V_s'+
ie_{2t}JV'_tV_s^2+e_{2t-2s}V'_tV_s'\rt]ds.
\end{aligned}
\]
Consider $Y_{2,a}=K^2Vy^o$. We have
\[
\begin{aligned}
\lb{Y3} Y_{2,a}=e_{1}K^2Vy^o=
\int_0^1\!\!dt\!\!\int_0^t(iJV_t^2-e_{2t}V'_t)(e_{2s}iJV_s^3-V_s'V_s)ds
\\
=\int_0^1dt\int_0^t\rt[-e_{2s}V_t^2V_s^3-iJV_t^2V_s'V_s
+ie_{2t-2s}JV'_tV_s^3+ e_{2t}V'_tV_s'V_s\rt]ds.
\end{aligned}
\]
Consider $Y_{2,b}=KVKy^o$. We have
\[
\begin{aligned}
\lb{Y4} Y_{2,b}=e_{1}KVKy^o=
\int_0^1\!\!dt\!\!\int_0^t(ie_{2t}JV_t^3-V'_tV_t)
(iJV_s^2-e_{2s}V_s')ds
\\
=\int_0^1dt\int_0^t\rt[e_{2t}V_t^3V_s^2-ie_{2t-2s}JV_t^3V_s'
-iJV'_tV_tV_s^2+e_{2s} V'_tV_tV_s'\rt]ds.
\end{aligned}
\]
\no 3) Consider the third term $X_3=(Ka)^3\p_0= K^3\p_0+O(\ve
e^{|\n|})$. From \er{31}, \er{ia2}  we have
\[
\begin{aligned}
\lb{Y5} & Y_{3,1}=e_{1}K^3\p_0=\!\!\int_0^1\!\!dt\!\!\int_0^tds
 \int_0^s\!\! f_{3,1}dr,
\\
 & f_{3,1}=(iJV_t^2-e_{2t}V'_t)(iJV_s^2-e_{2s}V_s')(iJV_r^2-e_{2r}V'_r),
\end{aligned}
\]
where
\[
\begin{aligned}
\lb{Y6}  f_{3,1}
=\big[-V_t^2V_s^2-iJe_{2s}V_t^2V_s'+e_{2t}iJV'_tV_s^2+e_{2t-2s}V_t'V'_s\big]
(iJV_r^2-e_{2r}V'_r)
\\
=-iJV_t^2 V_s^2V_r^2+e_{2r}V_t^2 V_s^2V'_r
 -e_{2s}V_t^2V_s'V_r^2+iJe_{2s-2r}V_t^2V_s'V'_r
 \\
 +e_{2t}V'_tV_s^2V_r^2-e_{2t-2r}iJV'_tV_s^2V'_r+e_{2t-2s}iJV_t'V'_sV_r^2-
 e_{2t+2r-2s}V_t'V'_sV'_r.
\end{aligned}
\]
Finally, collecting all these computations we obtain the asymptotics
of $\P$ and its trace $T$.

\begin{corollary} \lb{TasP}
 Let $(\l,v')\in \mB_v \ts\mH$ and $\ve={1\/2\l}$. Then the function
$\P=a_0\p a_0^{-1}$ satisfies
\[
\lb{asP1}
\begin{aligned}
\P(\l)=e^{-i\l J}\bigg(\1+\ve Y_1(\l)+\ve^2Y_2(\l)
+\ve^3Y_3(\l)\bigg)+e^{|\n|}O(\ve^4),
\end{aligned}
\]
\[
\lb{2tr1}
\begin{aligned}
\tes T=\Tr \p=\Tr \P=T_0+\ve \gt_1+\ve^2\gt_2+\ve^3\gt_3
+e^{|\n|}O(\ve^4),
\end{aligned}
\]
where $\gt_j=\Tr e^{-i\l J} Y_j, j\in \N_3$, as $|\l|\to \iy$,
uniformly on bounded subsets of $(\arg \l,v)\in [0,2\pi]\ts \mH$, and
\[
\lb{asP2}
\begin{aligned}
& Y_1=iJ\mV-\wh {V_\l'},\qqq Y_2=-Y_2^o+Y_2^1,
\qqq 
 Y_2^o=\!\!\int_0^1\!\!
V_t'V_tdt+\!\!\int_0^1\!\!dt\!\!\int_0^t\!\! V_t^2V_s^2dds,
 \\
 &   Y_2^1=iJ\wh {V^3_\l}+\!\!\int_0^1\!\!dt\!\!\int_0^t\!\!
\rt(e_{2t-2s}V'_tV_s'+ie_{2t}JV'_tV_s^2-ie_{2s}JV_t^2V_s'
\rt)ds,
\\
& Y_3=Y_{1,3}+Y_{3,1}+Y_{2,a}+Y_{2,b},
\end{aligned}
\]
where $Y_{1,3}, Y_{3,1}, Y_{2,a},Y_{2,b}$ are given by
\er{Y1}-\er{Y6}; $\wh {V_\l}=\int_0^1e_{2t}V_tdt$  and $e_t=e^{i\l tJ}$.
\end{corollary}

\subsection{Traces for $v'\in \mH$}

We compute traces, which will be used to study conformal
mapping.

\begin{lemma}  \lb{TT1}
Let $(\l,v')\in \mB \ts\mH$. Then the traces $\gt_j=\Tr e^{-i\l J} Y_j, j=1,2,3$
 are given by
\[
\lb{2tr2}
\begin{aligned}  \gt_1=2\|v\|^2\sin \l,
\end{aligned}
\]
\[
\lb{2tr3}
\begin{aligned}
& \gt_2=-2\cH_1\sin \l-\|v\|^4\cos \l-e^{i\l}C_{1,2}+\gt_{(2)},
\\
& \gt_{(2)}= e^{-i\l}T_{2}^++e^{i\l}T_{2}^-, \qq
T_{2}^+=\!\!\int_0^1\!\!\int_0^{s_1}\!\! e^{i2\l(s_1-s_2)}
{v'}^*(s_1) v'(s_2)ds,\qqq T_{2}^-=\wt T_{2}^+,\
\end{aligned}
\]
where $\cH_1=-i\lan v',v\ran$, \
$C_{1,2}=|\gc_{12}|^2-\|v_1\|^2\|v_2\|^2$, $\ \
\gc_{12}=\int_0^1v_1\ol v_2dx$ and
\[
\lb{2tr4}
\begin{aligned} \gt_3=2(\sin \l)\|v\|_4^4+\gt_{3,2}+\gt_{3,1},\qqq
\gt_{3,1}=\!\!\int_0^1\!\!dt\!\!\int_0^tds \!\! \int_0^s\!\!
F_{3,1}dx,
\end{aligned}
\]
\[
\lb{2tr5}
\begin{aligned}
\gt_{3,2}=\!\!\int_0^1\!\!dt\!\!\int_0^t\! i\Tr
e_{-1}Je_{2t-2s}\rt[V'_tV_s^3-V_t^3V_s'\rt]ds=e^{|\n|}O(\ve),
\\
F_{3,1}=\Tr e_{-1}iJ\rt[ e_{2s-2x}V_t^2V_s'V'_x
-e_{2t-2x}V'_tV_s^2V'_x+e_{2t-2s}V_t'V'_sV_x^2-V_t^2V_s^2V_x^2\rt].
\end{aligned}
\]
 Moreover, we have
 \[
 \lb{2tr6}
 \begin{aligned}
\gt_2-\wt \gt_2=-2iC_{12}\sin \l, \qqq \gt_3-\wt
\gt_3=\!\!\int_0^1\!\!dt\!\!\int_0^tds \!\! \int_0^s\!\! (F_{31}-\wt
F_{31})dx.
\end{aligned}
 \]

\end{lemma}

 \no {\bf Proof.}
1) Consider $\gt_1=\Tr e_{-1}Y_1$. From \er{215} and \er{217} we
obtain
$$
\begin{aligned}
\gt_1=\Tr e_{-1}(iJ\mV-\wh {V'}(\l))=\Tr e_{-1}iJ\mV
=\Tr (iJ\cos \l+\sin \l)\mV=\sin \l\Tr \mV=2\sin \l \|v\|^2.
\end{aligned}
$$
2) Consider $\gt_2=\Tr e_{-1}Y_2$ given by \er{asP2}. The trace of
the first term  $\Tr e_{-1}Y_2^o$ in\er{asP2} has the form
$$
\begin{aligned}
& \Tr
e_{-1}Y_2^o=  \Tr e_{-1}\rt( \!\!\int_0^1\!\!
V_t'V_tdt+\!\!\int_0^1\!\!dt\!\!\int_0^t\!\! V_t^2V_s^2dds \rt),
\\
& \Tr
e_{-1}\!\!\int_0^1\!\! V_t'V_tdt=e^{-i\l}\lan v,v'\ran+e^{i\l}\lan v',v\ran
=-2\cH_1\sin \l,
\\
&  \Tr  e_{-1}\!\!\int_0^1\!\!dt\!\!\int_0^t\!\!
V_t^2V_s^2ds=\Tr e_{-1}{\mV^2\/2}={e^{-i\l}\/2}\|v\|^4
+{e^{i\l}\/2}(\gb_1^2+\gb_2^2)=\|v\|^4\cos \l+e^{i\l}C_{1,2},
\end{aligned}
$$
where $\gb_1^2+\gb_2^2=\|v\|^4+2C_{1,2}$, and
$C_{1,2}=|\gc_{1,2}|^2-\|v_1\|^2\|v_2\|^2$.
Consider the trace of the second term  $\Tr e_{-1}Y_2^1$ in\er{asP2}.
Due to \er{215} the trace of the  last term of $Y_2$ from \er{asP2} have the form
$$
\begin{aligned}
& \Tr e_{-1}Y_2^1=\Tr
e_{-1}iJ\wh {V^3}+\Tr  e_{-1}\!\!\int_0^1\!\!dt\!\!\int_0^t\!\!
\rt(e_{2s-2t}V'_tV_s'-ie_{2s}JV_t^2V_s'-
ie_{2t}V'_tJV_s^2\rt)ds
\\
& = \Tr  \!\!\int_0^1\!\!dt\!\!\int_0^t\!\!
e_{-1}e_{2s-2t}V'_tV_s'ds
=e^{-i\l}T_2^{+}+e^{i\l}T_2^{-}.
\end{aligned}
$$
\no 3) Consider the more complicated case $\gt_3=\Tr
e_{-1}Y_3=\gt_{1,3}+\gt_{3,2}+\gt_{3,1}$, where
$$
\lb{ty123}
\begin{aligned}
\gt_{1,3}=\Tr e_{-1}Y_{1,3},\ \ \gt_{3,1}=\Tr e_{-1}Y_{3,1}, \ \ \gt_{3,2}=\Tr
e_{-1}(Y_{2,a}+Y_{2,b}),
\end{aligned}
$$
and $Y_{1,3}, Y_{3,1}, Y_{2,a}, Y_{2,b}$ are  given by
\er{Y1}-\er{Y6}.

Consider $\gt_{1,3}=\Tr e_{-1}Y_{1,3}$. From \er{213} and \er{215} we obtain
\[
\lb{ty13}
\begin{aligned}
\gt_{1,3}=\Tr e_{-1}Y_{1,3}= \Tr e_{-1}\rt(iJ\int_0^1
V^4_tdt-\int_0^1e^{i2t\l J}V_t'V_t^2dt\rt)=\Tr e_{-1}iJ\int_0^1
V^4_tdt
\\
=\int_0^1 \Tr \rt[i(\cos \l) JV_x^4+(\sin \l)
 V_x^4\rt]dx=(\sin \l)\int_0^1\Tr V_x^4dx=2(\sin \l)\|v\|_4^4.
\end{aligned}
\]
Consider $\gt_{3,1}=\Tr e_{-1}Y_{3,1}$. From \er{Y6} we obtain
\[
\lb{ty31}
\begin{aligned}
& \gt_{3,1}=\Tr e_{-1}Y_{3,1}= \!\!\int_0^1\!\!dt\!\!\int_0^tds \!\!
\int_0^s\!\! F_{3,1}dr,
\\
& F_{3,1}=\Tr e_{-1}\rt[ -iJV_t^2 V_s^2V_r^2+e_{2r}V_t^2 V_s^2V'_r
 -e_{2s}V_t^2V_s'V_r^2+iJe_{2s-2r}V_t^2V_s'V'_r
 \\
&  +e_{2t}V'_tV_s^2V_r^2-e_{2t-2k}iJV'_tV_s^2V'_r+e_{2t-2s}iJV_t'V'_sV_r^2-
 e_{2t+2r-2s}V_t'V'_sV'_r  \rt].
\end{aligned}
\]
From \er{213} and \er{215} we obtain
\[
\begin{aligned}
\lb{ty31q}
 F_{3,1}=\Tr e_{-1}iJ\rt[ e_{2s-2r}V_t^2V_s'V'_r
-e_{2t-2r}V'_tV_s^2V'_r+e_{2t-2s}V_t'V'_sV_r^2-V_t^2V_s^2V_r^2\rt].
\end{aligned}
\]
Consider the first term $\gt_{2,a}=\Tr e_{-1}  Y_{2,a} $ in
 $\gt_{3,2}=\gt_{2,a}+\gt_{2,b}$.
  From \er{Y3} and
\er{213}, \er{215} we obtain
\[
\lb{ty2a}
\begin{aligned}
& \gt_{2,a}=\!\!\int_0^1\!\!dt\!\!\int_0^t \!\! \Tr
e_{-1}\rt[-e_{2s}V_t^2V_s^3-iJV_t^2V_s'V_s +iJe_{2t-2s}V'_tV_s^3+
e_{2t}V'_tV_s'V_s\rt]ds
\\
& =i\!\!\int_0^1\!\!dt\!\!\int_0^t\! \Tr
e_{-1}J\rt[e_{2t-2s}V'_tV_s^3-V_t^2V_s'V_s \rt]ds.
\end{aligned}
\]
Consider  $\gt_{2,b}=\Tr e_{-1}  Y_{2,b}$. From \er{Y4}
and \er{213}, \er{215}  we
obtain
\[
\lb{ty2b}
\begin{aligned}
& \gt_{2,b}=\!\!\int_0^1\!\!dt\!\!\int_0^t \!\! \Tr e_{-1}
\rt[-e_{2t}V_t^3V_s^2-ie_{2t-2s}JV_t^3V_s'
-iJV'_tV_tV_s^2+e_{2s} V'_tV_tV_s'\rt]ds
\\
& =i\!\!\int_0^1\!\!dt\!\!\int_0^t \!\! \Tr
e_{-1}J\rt[V'_tV_tV_s^2     -e_{2t-2s}V_t^3V_s'\rt]ds.
\end{aligned}
\]
Then \er{ty2a}, \er{ty2b} give
\[
\lb{tyA}
\begin{aligned}
\gt_{3,2}=-i\Tr\!\!\int_0^1\!\!dt\!\!\int_0^t\!
 e_{-1}J\rt[V_t^2V_s'V_s+V'_tV_tV_s^2-e_{2t-2s}V'_tV_s^3
+  e_{2t-2s}V_t^3V_s'   \rt]ds
\\
=i\Tr\!\!\int_0^1\!\!dt\!\!\int_0^t\!
 e_{-1}J\rt[e_{2t-2s}V'_tV_s^3-e_{2t-2s}V_t^3V_s'   \rt]ds,
\end{aligned}
\]
since we have
$$
\begin{aligned}
& \Tr e_{-1}J\rt[ \!\!\int_0^1\!\!dt\!\!\int_0^t\!
V_t^2V_s'V_s +V'_tV_tV_s^2  \rt]ds
\\
&=\Tr e_{-1}J\rt[\!\!\int_0^1\!\!V_t^2dt\!\!\int_0^t\!
V_s'V_sds  +  \!\!\int_0^1\!\!V_s^2ds\!\!\int_s^1\!    V'_tV_tdt  \rt]
=\Tr e_{-1}J \!\!\int_0^1\!\!V_t^2dt\!\!\int_0^1\!
V_s'V_sds=0.
\end{aligned}
$$
Consider  the intergal $\gt_{3,2}$ in \er{tyA}. For the first
integral we have
$$
\begin{aligned}
& \int_0^t\! e_{2s}V_s^3ds=-\int_0^t\! i\ve J(e_{2s})'V_s^3ds =i\ve
J\rt[V_0^3-e_{2t}V_t^3+\int_0^t\! e_{2s}(V_s^3)'ds\rt],
\\
& \!\!\int_0^1\!\!dt\!\!\int_0^t \!\! \Tr
ie_{2t-1-2s}JV_t'(V_s)^3ds=\ve \!\!\int_0^1\!\!dt \Tr e_{2t-1}
V_t'\rt[V_0^3-e_{2t}V_t^3+\int_0^t\!
e_{2s}(V_s^3)'ds\rt]=e^{|\n|}O(\ve).
\end{aligned}
$$
Similar arguments imply
\[
\lb{tyB}
\begin{aligned}
\gt_{3,2}
=i\Tr\!\!\int_0^1\!\!dt\!\!\int_0^t\!
 e_{-1}J\rt[e_{2t-2s}V'_tV_s^3-e_{2t-2s}V_t^3V_s'   \rt]ds=e^{|\n|}O(\ve).
\end{aligned}
\]
Finally, collecting all these computations  we obtain $\gt_1, \gt_2,
\gt_3$ in \er{2tr2}-\er{2tr5}.
Moreover, using \er{2tr2}-\er{2tr5} we obtain \er{2tr6}.
\BBox

We discuss $ \gt_2(\l),
\gt_3(\l)$ as $\Im \l\to +\iy$, since the function $\gt_1=2\|v\|^2\sin \l$
has a good form.

\begin{lemma}  \lb{Tt2}
Let  $\cH_0=\|v\|^2, \ \cH_1=-i\lan v',v\ran$ and $
\cH_2=\|v'\|^2+\|v\|_4^4$ for $v'\in\mH$, where $\lan
\cdot,\cdot\ran$ is the scalar product in $\mH$. Let $\l\in \mD_r,
|\l|\to \iy$ for any $r>0$. Then the traces $T$ and $\gt_2, \gt_3$
have asymptotics uniformly on bounded subsets of $ \mH$:
\[
\lb{tt2}
\begin{aligned} \gt_2(\l)=\big[-i\cH_1-{\cH_0^2\/2}+
\ve i\|v'\|^2+ o(\ve)\big]e^{-i\l},
\end{aligned}
\]
\[
\lb{tt3}
\begin{aligned}
\gt_3(\l)= \Big[ i\|v\|_4^4+ \cH_0 \cH_1-i{\cH_0^3\/3!}+o(\ve)\Big]e^{-i\l},
\end{aligned}
\]
\[
\lb{tt4}
\begin{aligned}
&   T(\l)=\Big[1+\ve i\cH_0+\ve^2\Big(-i\cH_1-{\cH_0^2\/2} \Big)
+\ve^3\Big(i \cH_2+ \cH_0 \cH_1-i{\cH_0^3\/3!}\Big)+o(\ve^3)\Big]e^{-i\l}
\\
&   =\exp \Big[-i\Big(\l-\cH_0\ve -\cH_1 \ve^2-\cH_2\ve^3+o(\ve^3)\Big)
\Big].
\end{aligned}
\]

\end{lemma}

 \no {\bf Proof.}  Consider $\gt_2$ defined in \er{2tr3}.
We need to discuss only $T_{2}^+$ in \er{2tr3}, since all other are
simple.
 Due to \er{TA-1} the term $T_{2}^+$ given by \er{2tr3} satisfies
$$
\tes T_{2}^+=\!\!\int_0^1\!\!\int_0^{s_1}\!\! e^{i2\l(s_1-s_2)}
{v'}^*(s_1) v'(s_2)ds ={i\/2\l }\lt(\int_0^1
{v^*}'(t){v'}(t)dt+o(1)\rt)={i\/2\l }(\|v'\|^2+o(1)).
 $$
Consider $\gt_3$ in \er{2tr4}. The first term has the form
 \[
\lb{ttq2} \Tr e_{-1}iJ\int_0^1V^4_xdx= 2(\sin
\l)\|v\|_4^4=i\|v\|_4^4(e^{-i\l}-e^{i\l}).
\]
Consider the next term  $G=\!\!\int_0^1\!\!dt\int_0^t\! F_{3,2}ds$
in \er{2tr4}. From  \er{2tr4} we have
\[
\lb{ttq3}
\begin{aligned}
G=-i\!\!\int_0^1\!\!dt\!\!\int_0^t\! \Tr e_{-1}J\rt[V_t^2V_s'V_s
+V'_tV_tV_s^2 -e_{2t-2s}V'_tV_s^3
+  e_{2t-2s}V_t^3V_s'     \rt]ds
\\
=G_{o}+O(\ve e^{-i\l}),
  \qqq
G_{o}=-i\!\!\int_0^1\!\!dt\!\!\int_0^t\! \Tr
e_{-1}J\rt[V_t^2V_s'V_s+V'_tV_tV_s^2
  \rt]ds,
\end{aligned}
\]
since due to \er{TA-1} we have
$$
\!\!\int_0^1\!\!dt\!\!\int_0^t\! \Tr e_{-1}J\rt[-e_{2t-2s}V'_tV_s^3
+%
  e_{2t-2s}V_t^3V_s'     \rt]ds=e^{\n}O(\ve).
$$
We compute asymptotic of $G_{o}$.
We have
\[
\lb{G2}
\begin{aligned}
G_{o}=-i\!\!\int_0^1\!\!dt\!\!\int_0^t\! \Tr e_{-1}J\rt[V_t^2V_s'V_s+V'_tV_tV_s^2 \rt]ds
=-i\!\!\int_0^1\!\!dt\!\!\int_0^1\! \Tr e_{-1}J
V_t^2V_s'V_sds
\\
=-ie^{-i\l}\!\!\int_0^1\! \|v\|^2 v_s'v_s^*ds+O(e^{-\n})
=-ie^{-i\l}\|v\|^2 \lan v',v\ran+O(e^{-\n}).
\end{aligned}
\]
Consider $\gt_{3,1}=\Tr e_{-1}Y_{3,1}$ in \er{2tr4}. From \er{TA-2}
 we obtain
$$
\begin{aligned}
& \gt_{3,1}=\int_G
i\Tr e_{-1}J\rt[ -V_t^2 V_s^2V_x^2+
 e_{2s-2x}V_t^2V_s'V'_x-e_{2t-2x}V'_tV_s^2V'_x+
 e_{2t-2s}V_t'V'_sV_x^2\rt]dsdtdx
 \\
& =G_{3,1}+e^{-i\l}O(\ve),\qqq
 G_{3,1}=-\!\!\int_\cG\!\! i\Tr e_{-1}JV_t^2 V_s^2V_x^2dsdtdx,
\end{aligned}
$$
where  $\cG=\{0<x<t<s<1\}$,   since due to \er{ab2} we obtain
\[
\!\!\int_\cG\!\! i\Tr e_{-1}J\rt[
 e_{2s-2x}V_t^2V_s'V'_x-e_{2t-2x}V'_tV_s^2V'_x+
 e_{2t-2s}V_t'V'_sV_x^2\rt]dsdtdx=O(\ve e^{-\n}).
\]
Consider $G_{3,1}$. Due to the identity $V_t^2=\ma |v_t|^2 & 0\\
0& v_tv_t^*\am$, we deduce that
$$
\begin{aligned}
& G_{3,1}
 = -i\!\!\int_\cG\!\! e^{-i\l}|v(t)|^2 |v(s)|^2|v(x)|^2dsdtdx
+O(e^{-\n})=-ie^{-i\l}{\|v\|^6\/3!}+O(e^{-\n}).
\end{aligned}
$$
Finally, collecting all these computations of $\gt_1, \gt_2, \gt_3$
we obtain the first asymptotics in \er{tt4}. In order to show the
second one in \er{tt4} we need a simple decomposition for $a,b,c,
\ve \in \C$:
\[
\lb{eea}
\begin{aligned}
e^{i(a\ve+b \ve^2+c\ve^3)}=
1+ia\ve+\ve^2\Big(ib-{a^2\/2}\Big)+\ve^3\Big(-i{a^3\/3!}-ab+ic\Big)+O(\ve^4),
\qqq |\ve |<1.
\end{aligned}
\]
We have the  first asymptotics in \er{tt4} and via \er{eea} we can
rewrite it in the needed form. Indeed assume that
 $$
 T=e^{-i\l+i(a\ve+b \ve^2+c\ve^3)+o(\ve^3)}.
 $$
Then \er{eea} gives the asymptotics in the form $e^{-i\l}[1+A_1\ve+
A_2\ve^2 + A_3\ve^3+ o(\ve^3)]$, where the coefficients $A_1,A_2,
A_3$ are given by \er{tt4}. From here and  \er{eea} we obtain $
a=\cH_0,\ b=\cH_, \ c=\cH_2$, which yields the second asymptotics in
\er{tt4}. \BBox


\section {The branch points of  the Riemann surface   \lb{Sec4}}
\setcounter{equation}{0}

We discuss asymptotics of multipliers,  Lyapunov functions and
discriminants.

 \no {\bf Proof of Theorem \ref{T2}.} We show \er{a1} for
$\l\in \ol\C_+$, the proof for $\l\in \ol\C_-$ is similar. From
\er{asm}  we have $\t_3(\l)=e^{-i\l+o(1)}$ and
$\t_j(\l)=e^{i\l+o(1)}, j=1,2$ as $|\l|\to \iy$, where $|\l-\pi
n|\ge \d, $ for all $n\in \Z$ and for each $\d\in (0,{1\/2})$. From
\er{36} we have
\[
\lb{asP}
\begin{aligned}
\P(\l)= e^{-i\l J}\rt[\1+\ve
 \big(iJ\mV-\wh {V'}(\l)\big)\rt]
 +O(\ve^2e^\n)=e^{-i(\l J-\ve J\mV)}+o(\ve e^\n),
 \end{aligned}
\]
since $e^{-i\l J}\wh {V'}(\l)=o(e^\n)$ and  the matrices $J\mV=\mV
J$. Consider the matrix $J\l-\ve J\mV$. Its eigenvalues are given by
$ \l-\ve \gb_1,\ -\l+\ve \gb_2,\ -\l+\ve \gb_3. $ Using the standard
arguments from the perturbation theory for matrices (see \cite{Ka95},
p.291]), we obtain that \er{asP} gives asymptotics \er{a1} for
$\t_3$. In order to determine asymptotics of $\t_1, \t_2$ we
consider $\P(\l)^{-1}$ with the eigenvalues $\t_1^{-1}, \t_2^{-1}$.
From \er{3} and  $\o=\1-\ve^2V_0^2, \ \ \ve={1\/2\l}$   we get
\[
\lb{Pp1} \tes \P^{-1}=a_0\p^{-1} a_0^{-1} =a_0J \wt \p J
a_0^{-1}=\a_0 Ja_0\wt\P(\l)a_0^{-1}Ja_0^{-1} =\o J\wt\P(\l)J\o^{-1}.
\]
From \er{asP} and \er{Pp1}  we obtain
\[
\lb{asPz}
\begin{aligned}
& \wt\P(\l)= e^{i(\l J-\ve J\mV)}+o(\ve e^\n),
\\
& \P(\l)^{-1}=\o J\wt\P(\l)J\o^{-1}=e^{i(\l J-\ve J\mV)}+o(\ve e^\n).
 \end{aligned}
\]
Thus using above arguments for $\t_3$ we obtain asymptotics of
$\t_1, \t_2$. Then \er{a1} implies  asymptotics of
$\D_j={1\/2}(\t_j+\t_j^{-1})$ in    \er{a2}. We show \er{a3}. From
\er{a1}, \er{rqm}  and for $\t_j=e^{i\f_j}, j=1,2,3$ we have
$$
\begin{aligned}
\tes
\gD=\sin^2{\f_1-\f_2\/2}\sin^2{\f_1-\f_3\/2}\sin^2{\f_2-\f_3\/2}
 =\sin^2(\l-k_{31}) \sin^2(\l-k_{32})
 \sin^2{\f_2-\f_1\/2},
 \end{aligned}
$$
which yield \er{a3}. \ \BBox

 \begin{lemma}  \lb{Tgf}
Let $\|v'\|\le r<m$ for some $r>0$ and $\ve={1\/2\l}$. Then

\no i) The function $\gf=\cT_--\sin \l$, where $\cT_-={1\/2i}(T-\wt T)$, has asymptotics
 \[
 \lb{gf1}
 \gf(\l)=\ve^2(\gb_1\gb_2+o(1))\sin \big[\l-{\ve\/2}(\gb_3+o(1))\big]
 \qquad\as \qq |\l|\to \iy.
 \]
ii) For $m\in \N$ large enough
$\gf (\l)$ has exactly $2m-1$ roots, counted
with multiplicities, in the disc $\dD_{\pi m+{\pi\/2}}$ and
 exactly one simple real root $b_{n}$ in each disc $\dD_{\vs}(\pi n), n>m, \vs\in (0,{1\/4})$,
 where the disc  $\dD_r(\l_o)=\{|\l-\l_o|<r\}$ for $r>0, \l_o\in \C$. There are no other roots.

\no iii)  The zero  $b_n\in \dD_{\vs}(\pi n)$ of  $\gf$  has
asymptotics 
\[
\lb{gf2} \tes
 b_n=\pi n+{\gb_3+o(1)\/4\pi n}\qqq  \as\qq n\to \pm \iy.
\]
Asymptotics \er{gf1}, \er{gf2} hold true   uniformly on bounded subsets of $\mH$.

\end{lemma}

 \no {\bf Proof.} i) Let  $\t_3(\l)=e^{-ik_3(\l)}$. From \er{a1} we obtain
 $$
 \begin{aligned}
e^{i\l}-\t_j(\l)=e^{i\l}[1-e^{-i\ve(\gb_j+o(1)}]=i\ve[\gb_j+o(1)]e^{i\l},
\qq j=1,2,
\\
\tes e^{i\l}-\t_3(\l)=    e^{i\l}-e^{-ik_3}=2ie^{i(\l-k_3)/2}\sin {\l+k_3\/2}=
(2i+o(\ve))\sin [\l-{\ve\/2}(\gb_3+o(1))].
\end{aligned}
 $$
 This and the identity $\gf={e^{-i2\l}\/2i}  D(e^{i\l},\l)$ imply
 $$
 \begin{aligned}
 \tes
 \gf=-{e^{-i2\l}\/2i}(e^{i\l}-\t_1(\l))(e^{i\l}-\t_2(\l))(e^{i\l}-\t_3(\l))
 =\ve^2(\gb_1\gb_2+o(1))\sin [\l-{\ve\/2}(\gb_3+o(1))].
\end{aligned}
 $$
 ii) The proof is standart and follows from asymptotics \er{gf1}. Note that
 the function $\gf$ is real on the real line and then it has a real zero
 $b_n$ in each disc $\dD_{\vs}(\pi n), |n|>m$.

\no iii)  From i) and ii) we have $\sin [\l-{\ve\/2}(\gb_3+o(1))]=0$ for $\l=b_n$,
which yields $\l-{\ve\/2}(\gb_3+o(1))=\pi n$
 and then \er{gf2}.
 \BBox

 \no {\bf Proof of Theorem \ref{T3}.} i) Introduce the contours
$
 C_m=\{\l: |\l|=\pi (m+{1\/2})\}$ and $ C_n(\vs)=\{\l:|\l-\pi n|=\vs\}$
for some $\vs\in (0,{1\/4})$. Let $m_1>m$ be another integer and let
$\l$ belong to the contours $C_{m}, C_{m_1}, C_n(\vs)$, where $m\le
n\le m_1$. From \er{a1} we have for $|\l|\to \iy$
 \[
\lb{ask2x}
\begin{aligned}
\tes \gD(\l)={1\/(4\l)^2}(\gb_o+w(\l))\vr_1(\l)\vr_2(\l),\qqq
w(\l)=o(1),
\\
\tes \vr_j=\sin^2 (\l-\f_j), \qqq
\f_j={1\/4\l}(\gb_3+\gb_j+o(1)),\qq j=1,2,
\end{aligned}
\]
uniformly in $\arg \l\in [0,2\pi]$.
The function $\sin(\l-\f_j)$ and $\vr_j$  satisfy
 $$
 \begin{aligned}
& \tes \sin(\l-\f_j)-\sin \l=-2\sin{\f_j\/2}\cos(\l-{\f_j\/2}),
\\
& |\sin(\l-\f_j)-\sin \l|\le |\f_j|e^{|\Im \l|+|{|\f_j|\/2}}, \qqq
|\vr_j(\l)-\sin^2 \l|\le 2|\f_j|  e^{2|\Im \l|+||\f_j|},
\end{aligned}
$$
and
 $$
 \begin{aligned}
\tes
\big| (\sin ^4\l-\vr_1(\l)\vr_2(\l)\big|\le |\sin ^2\l| |\sin ^2\l-\vr_1(\l)|+
|\vr_1(\l)||\sin ^2\l-\vr_2|
\\
\le 2|\f_1|  e^{4|\Im \l|+||\f_1|}+2|\f_2|  e^{4|\Im \l|+2|\f_1|+|\f_2|}\le  A_1 e^{4|\Im \l|},
\end{aligned}
$$
where $A_1=2(|\f_1+|\f_2|)e^{2|\f_1|+|\f_2|}$, and
$$
\big|w(\l)\vr_1(\l)\vr_2(\l)\big|\le |a|e^{4|\Im \l|+||\f_1|+|\f_2|}=\gb_oA_2e^{4|\Im \l|},
\qq A_2={|a|\/\gb_o}e^{||\f_1|+|\f_2|}.
$$
This gives for $\gD$ and $ \gD_o(\l):=\gb_o {\sin ^4\l\/(4\l)^2}$ the following
 $$
 \begin{aligned}
\tes |\gD(\l)-\gD_o(\l)|={1\/|4\l|^2}
|\gb_o \big[\sin ^4\l-\vr_1(\l)\vr_2(\l)\big]-w(\l)\vr_1(\l)\vr_2(\l)|
\\
\tes \le {\gb_o\/|4\l|^2}\big| (\sin
^4\l-\vr_1(\l)\vr_2(\l)\big|+{|w(\l)|\/|4\l|^2}
\big|\vr_1(\l)\vr_2(\l)\big|\le  {\gb_o\/|4\l|^2}e^{4|\Im \l|}A,
\end{aligned}
$$
where $A=A_1+A_2$. Then using the estimate $e^{|\Im \l|}\le
{2\/(1-e^{-2\vs})}|\sin \l|$   from \er{FuT1} we obtain
$$
|\gD(\l)-\gD_o(\l)|\le A w^4 {\gb_o\/|4\l|^2}|\sin^4 \l|\le {1\/2}|\gD_o(\l)|,
$$
on all contours $C_{m}, C_{m_1}, C_n(\vs)$ for $m<n<m_1$,
since $A_1=o(1)$ and $A_2=o(1)$ as $|\l|\to \iy$, uniformly
in $\arg \l\in [0,2\pi]$.
 Hence, by Rouch\'e's theorem, $\gD$ has as many roots,
counted with multiplicities, as $\gD_o$ in each of the bounded domains and
the remaining unbounded region. Since $\gD_o$ has only four  roots in each disc
$\{ |\l-\pi n|<\vs \}, |n|>m$ and since $m_1>m$ can be chosen arbitrarily large,
the point i) of Lemma follows.

 ii) Let $\l\in I_n=[\pi(n-1)+\vs, \pi n-\vs]$ and $n\to \pm\iy$. From \er{ask2x} we have
 $$
 \begin{aligned} (4\l)^2\gD(\l)=(\gb_1-\gb_2+o(1))^2\sin^4(\l+o(1))
 =(\gb_o+o(1))\sin^4(\l+o(1)).
\end{aligned}
 $$
 Then the function $\gD$ is  positive on $I_n$ for $n$ large enough,
 since it  is real on the real line.

 iii) Due to i) the function $\gD$ has 4 zeros $\l_{n,j}^\pm, j=1,2 $ in the disc $\dD_\vs(\pi
 n)$ for large $n\in \Z$ and small $\vs>0$.
 From \er{a1} and $\t_3=\t_j$ for $j=1$ or $j=2,$ we obtain for $\l=\l_{n,j}^\pm\in \dD_\vs(\pi  n)$:
$$
\tes e^{-i(\l-\ve \gb_3+o(\ve))}= e^{i(\l-\ve \gb_j+o(\ve)} ,\qq
$$
which yields $2\l -\ve (\gb_3+\gb_j)+o(\ve)=2\pi n$ and then we have
\er{ask3}.

We show that they are real. Assume that all four branch points in
$\l_{n,j}^\pm\in \dD_\vs(\pi n), j=1,2 $ are complex. Then the
Lyapunov functions $\D_j, j\in \N_3$ are real on $g_n(\vs)=(\pi
n-\vs,\pi n+\vs) $ and have the maximum $=1$ ( or minimum $=-1$, see
Fig. 2)  and the functions $\D_3-\D_s, s=1,2$ have simple zeros on
$g_n(\vs)$, which gives the contradiction.
 \BBox

\section {Lyapunov functions}
\setcounter{equation}{0}

\subsection{The case $v'\in \mH$   \lb{Sec5}}

Recall that $\P=a_0^{-1}\p a_0$ and $\wt a=a$.
In order to determine the asymptotics of the
Lyapunov function we need the following modification.

\begin{lemma}  \lb{TLf1}
Let $v'\in \mH$ and  $\o=(\1_3-\ve^2V_0^2)^{-1}$, where $\ve={1\/2\l}$.
Then the function  $\P=a_0^{-1}\p a_o$ satisfies
\[
\lb{51} \tes \P^{-1}=\o^{-1}J\wt\P J\o, \qq \l\in \C,
\]
\[
\lb{52}  J\wt\P J= J\wt Y Je^{i\l J},\qq
J\wt YJ=\1+\ve J\wt Y_1J+\ve^2 J\wt Y_2J+O(\ve^3),
\]
uniformly on bounded subsets of $\{|\Im \l|\le C\}\mH$ for some $C>0$,
where  $Y_1, Y_2$ are given by \er{asP1} and
\[
\lb{53}
\begin{aligned}
 J\wt Y_1J=-Y_1,  \qq
J\wt Y_2J=-\wt Y_2^o+\wt Y_2^1, \qq
\wt Y_2^o= \!\!\int_0^1\!\!V_xV_x'dx+\!\!\int_0^1\!\!dt\!\!\int_0^t\!\!
V_s^2V_t^2ds,
\\
J\wt Y_2^1J=-iJ\wh {V_\l^3}+\!\!\int_0^1\!\!dt\!\!\int_0^t\!\!
\rt(e_{2s-2t}V'_sV_t'-ie_{2t}JV_s^2V'_t+ie_{2s}JV_s'V_t^2\rt)ds.
\end{aligned}
\]
\end{lemma}
 \no {\bf Proof.}
  Substituting the identities
$
\p^{-1}=J\wt \p J $ and $\wt \p=a_0^{-1}\wt\P a_0
$
into $\P^{-1}=a_0^{-1}\p^{-1}a_o$  and using the identity $a_0^{-1}J a_0^{-1}=\o J$
since  $JV J=-V$,   we obtain
\[
\lb{54}
\tes \P^{-1}=a_0^{-1}J a_0^{-1}\wt\P a_0
Ja_0=\o^{-1}J\wt\P J\o,\ \
\]
which  yields \er{51}. Consider $\P^{-1}=\o^{-1}J\wt\P J\o$. From \er{asP1} we have
$$
\begin{aligned}
J\wt\P J=J  \rt(\1+\ve \wt Y_1+\ve^2 \wt Y_2+
O(\ve^3)\rt) e^{i\l J}J
=e^{i\l J}+J\rt(\ve \wt Y_1+\ve^2 \wt Y_2
\rt) e^{i\l J}J+O(\ve^3).
\end{aligned}
$$
Consider $J  \wt Y_1 J$.  From $ Y_1=iJ\mV-\wh {V_\l'}$ we obtain
$\wt Y_1=-iJ\mV-\wh {V_\l'}$, which yields
$$
\begin{aligned}
\lb{}  J \wt Y_1J= J (-iJ\mV-\wh {V_\l'}) J
= (-iJ\mV+\wh {V_\l'})=-Y_1,
\end{aligned}
$$
which yields \er{53}.
Consider $J  \wt Y_2 J$.  From \er{asP2} we obtain
$$
\begin{aligned}
\lb{}  J \wt Y_2J=-\wt Y_2^o+\wt Y_2^1,\qq \wt Y_2^o=
 \!\!\int_0^1\!\!V_xV_x'dx+\!\!\int_0^1\!\!dt\!\!\int_0^t\!\!
 V_s^2V_t^2ds,
\\
J\wt Y_2^1J=J \rt\{  -i\wh {V_\l^3}
+\!\!\int_0^1\!\!dt\!\!\int_0^t\!\!
\rt(e_{2s-2t}V'_sV_t'-ie_{2t}V_s^2V'_tJ+ie_{2s}V_s'V_t^2J  \rt\}
\\
= -iJ\wh {V_\l^3}+\!\!\int_0^1\!\!dt\!\!\int_0^t\!\!
\rt(e_{2s-2t}V'_sV_t'-ie_{2t}JV_s^2V'_t+ie_{2s}JV_s'V_t^2\rt)ds,
\end{aligned}
$$
which yields \er{53}. \BBox

Define a $\M^3$-valued function
$\gL={1\/2}(\P+\P^{-1})$ with eigenvalues
$\D_j={1\/2}(\t_j+\t_j^{-1}),j\in \N_3$.

\begin{lemma}  \lb{TLf2}
Let $v'\in \mH$ and $\o=(\1_3-\ve^2 V_0^2)^{-1}$. Then the function $\gL={1\/2}(\P+\P^{-1})$ satisfies
\[
\lb{L1} \gL={1\/2}(\P+\P^{-1})={1\/2}(\P+\o^{-1}J\wt\P J\o),
\]
\[
\lb{L2} \gL=\cos \l+ (\sin \l) \mV\ve +\gL_2\ve ^2+O(\ve^3)
\]
\[
\begin{aligned}
\lb{L5} \gL=\cos \Big(\l-\mV\ve +iJ A\ve ^2,
\Big)+\ve^2O\Big(|\ve|+|\wh {V'_\l}|\Big),
\end{aligned}
\]
as $\ve\to 0$,
uniformly on bounded subsets of $\{|\Im \l|\le C\}\ts \mH$, where
\[
\lb{L3} \gL_2={1\/2}\Big(\cos \l (\mV^2+{\wh {V'_\l}}^2)-iJ\sin \l
(A-A_2) +iJe^{-i\l J}[\wh {V'_\l},\mV]\Big),
\]
\[
\lb{L4}
\begin{aligned}
& A=\int_0^1\![V',V]dx+\int_0^1\! dt\int_0^t[V_t^2, V_s^2]_cds,
\\
& A_2=\!\!\int_0^1\!\!dt\!\!\int_0^t
(F(t,s)-F(s,t))ds,\qqq F(t,s)=e_{2t-2s}V'_tV_s'.
\end{aligned}
\]
\end{lemma}
\no {\bf Proof.} Identity \er{51} gives \er{L1}.   From \er{asP1}
and \er{asP1} we have asymptotics
$$
\begin{aligned}
\P=e^{-i\l J}(\1+\ve Y_1+\ve^2Y_2)+O(\ve^3),
\\
\o^{-1}J\wt\P J\o
=(\1-\ve Y_1+\ve^2 J\wt Y_2J)e^{i\l J}+O(\ve^3),
\end{aligned}
$$
as $\l\to \iy$. Then from \er{L1} we obtain
$$
\begin{aligned}
& \gL-\1_3\cos \l={\ve\/2}\Big(e^{-i\l J}(Y_1+\ve Y_2)+ (-Y_1+\ve J\wt Y_2J)e^{i\l J} \Big)+O(\ve^3) =\gL_1
\ve+\gL_2\ve^2+O(\ve^3),
\\
&  \gL_1={1\/2}\Big(e^{-i\l J}Y_1-Y_1e^{i\l J}\Big),\qqq
\gL_2={1\/2}\Big(e^{-i\l J}Y_2+J\wt Y_2 J e^{i\l J}\Big).
\end{aligned}
$$
Consider $\gL_1$.  From $\wh {V'}e^{i\l J}=e^{-i\l J}\wh {V'}$
and $Y_1=iJ\mV-\wh {V'}$ we obtain
\[
\begin{aligned}
\lb{} \gL_1={1\/2}\Big(e^{-i\l J}Y_1-Y_1 e^{i\l J}J\Big)
={1\/2}\Big(e^{-i\l J}iJ\mV-iJ\mV e^{i\l J}\Big)= \mV \sin \l.
\end{aligned}
\]
Consider $\gL_2$.  From \er{asP1}, \er{53} we obtain
$$
\begin{aligned}
 2\gL_2=e^{-i\l J}Y_2+J\wt Y_2 J e^{i\l J}
=e^{-i\l J}(-Y_2^o+Y_2^1)+(-{Y_2^o}^*+J\wt Y_2^1J) e^{i\l J}=-\gL_2^o+\gL_2^1,
\end{aligned}
$$
where $\gL_2^1=e^{-i\l J}Y_2^1+J\wt Y_2^1 Je^{i\l J}$ and $e^{i\l J}=\cos\l+iJ\sin \l$ gives
$$
\begin{aligned}
\gL_2^o=e^{-i\l J}Y_2^o+{Y_2^o}^* e^{i\l J}
=(Y_2^o+{Y_2^o}^*)\cos \l-iJ(Y_2^o-{Y_2^o}^*)\sin \l,
\\
Y_2^o+{Y_2^o}^*=\int_0^1\!\!(V'V+VV')dx+\int_0^1\!\!dt\int_0^t\!
(V_t^2V_s^2+V_s^2V_t^2)ds=\Big(\int_0^1\!\!V^2dx\Big)^2=\mV^2,
\end{aligned}
$$
since $\int_0^1(V^2)'dx=0$. Using $\int_0^1(VV'+V'V)dx=0$ we obtain
$$
\begin{aligned}
Y_2^o-{Y_2^o}^*=\!\int_0^1\!\! (V'V-VV')dx+\!\int_0^1\!\!\int_0^t\!
[V_t^2,V_s^2]_cdtds=2\int_0^1\!\!
V'Vdx+\!\int_0^1\!\!\int_0^t\! [V_t^2,V_s^2]_cdtds.
\end{aligned}
$$

Consider the next term $\gL_2^1$. From \er{asP1}, \er{53} we obtain
$$
\begin{aligned}
\gL_2^1=e^{-i\l J}Y_2^1+J\wt Y_2^1 Je^{i\l J}=
\rt(e^{-i\l J}iJ\wh {V_\l^3}   -iJ\wh {V_\l^3}e^{i\l J}\rt)+F_o+F_1=F_o+F_1,
\end{aligned}
$$
where
$$
\begin{aligned}
& F_o=\!\!\int_0^1\!\!dt\!\!\int_0^t (e_{-1}F(t,s)+
e_{1}F(s,t))ds,\qq F(t,s)=e_{2t-2s}V'_tV_s',
\\
& F_1=iJ\!\!\int_0^1\!\!dt\!\!\int_0^t\rt\{e^{-i\l J}
\rt(e_{2t}V'_tV_s^2-e_{2s}V_t^2V_s' \rt) +
\rt(-e_{2t}V_s^2V'_t+e_{2s}V_s'V_t^2 \rt)e^{i\l J} \rt\}ds,
\end{aligned}
$$
and using  $e^{i\l J}=\cos\l+iJ\sin \l$ we have
$$
\begin{aligned}
& F_o=\cos \l \!\!\int_0^1\!\!dt\!\!\int_0^1
F(t,s)ds+i(\sin\l)J    \!\!\int_0^1\!\!dt\!\!\int_0^t
(F(t,s)-F(s,t))ds=
\\
& \cos \l (\wh {V_\l'})^2+i(\sin\l)J    \!\!\int_0^1\!\!dt\!\!\int_0^t
(F(t,s)-F(s,t))ds.
\end{aligned}
$$
Consider the second term $F_1$ and we have
$$
\begin{aligned}F_1=iJe^{-i\l J}\!\!\int_0^1\!\!dt\!\!\int_0^t\rt(
e_{2t}V'_tV_s^2+e_{2s}V_s'V_t^2-e_{2s}V_t^2V_s'
-e_{2t}V_s^2V'_t
\\
=iJe^{-i\l J}\!\!\int_0^1\!\!\int_0^1\rt(
e_{2t}V'_tV_s^2-e_{2s}V_t^2V_s'\rt)ds=
iJe^{-i\l J}\rt(\wh {V'_\l}\mV-\mV \wh {V'_\l}\rt) .
\end{aligned}
$$
\BBox

Consider the asymptotics of the Lyapunov functions $\D_j, j\in \N_3$
as $\l\to \iy, |\Im \l|<C$.

\begin{corollary}  \lb{TLf3}
Let $v'\in \mH, C>0$. Then the functions $\D_j, j\in \N_3$
have  asymptotics:
\[
\begin{aligned}
\lb{7} \D_j(\l)=\cos \Big(\l-\ve \gb_j+\gr_j \ve^2\Big)
+\ve^2O\Big(|\ve|+|\wh {V'_\l}|\Big) \qq \as \qq 
|\l|\to \iy,\ |\Im \l|\le C,
\end{aligned}
\]
uniformly on bounded subsets of $\{|\Im \l|\le C\}\ts \mH$, where $\gr_1,\gr_2\in\R$ and $\gr_1+\gr_2+\gr_3=0$.

\end{corollary}
\no {\bf Proof.} From \er{asP1} and \er{asP1} we have asymptotics
$$
\begin{aligned}
\gL=\cos (\l \1-\ve S(\ve))+\ve^2O\Big(|\ve|+|\wh {V'_\l}|\Big),\qq S=\mV+\ve iJ A,
\end{aligned}
$$
where the matrices $\mV$ and $iJ A$ are self-adjoint. From \er{} we have $A=R+\cP$, where
$$
\begin{aligned}
R=\int_0^1\![V',V]dx= \ma R_1 & 0 \\ 0 & R_2\am,\qq
R_1=2\lan v',v\ran=2i\cH_1,
\\
\qq R_2=\int_0^1\! (v' v^*- v{v'}^*  )dx=2\int_0^1\! v {v'}^*dx,
\qq
\Tr R_2=2\lan v,v'\ran,
\end{aligned}
$$
and
\[
\lb{d1}
\begin{aligned}
& \cP=\int_0^1\! dt\int_0^t[V_t^2, V_s^2]ds= \ma 0 & 0 \\ 0 & \cP_2\am,\qq
\cP_2=\int_0^1\! dt\int_0^t[v(t)v^*(t),v(s)v^*(s)]ds,
\\
& A=R_1\os X, \qq X=R_2+\cP_2, \qq \Tr \cP_2=0,
\end{aligned}
\]
Let  $s_j, j=1,2,3$ be the eigenvalues  of $S(\ve)$,
which  are real for real $\ve$.
Since the matrices $S, \mV, R, \cP$ have the block forms,
we obtain $s_3(\ve)=\gb_3+\gr_3 \ve$, where $\gr_3=-2\cH_1$.
We compute  the eigenvalues of the $2\ts 2$ matrix $iX$, which  are zeros of the equation
$$
0=\det (iX-z)=z^2-z\Tr iX-\det X=z^2-z2\cH_1-\det X,
$$
since $\Tr iX=\Tr R_2=2i\lan v,v'\ran=2\cH_1$.
The perturbation theory for matrices
(see \cite{Ka95}) implies $s_j(\ve)=\gb_j+\gr_j \ve+O(\ve^2)$, which yields
\er{7}, where $\gr_1+\gr_2+\gr_3=0$  since $\Tr JA=0$.
\BBox


\section {Conformal mapping and the constants of motion \lb{Sec6}}
\setcounter{equation}{0}

\subsection{Conformal mapping}
Consider functions from the subharmonic  counterpart $\cS\cC$ of
the Cartright class  of the entire functions given by
 $$
 \cS\cC
 =\ca q: {\rm a \ real\ function \ }q(\l){\rm \  is \
subharmonic\  in \ \C \  and \ harmonic\  outside} \ \R,
 \\
 q(\ol \l)=q(\l),\ \l\in\C,
\\
 \int_{\R}{q_+(t)\/1+t^2}dt<+\iy,\ \
 \ a=\lim\sup_{\l\to\iy}{q(\l)\/|\l|}<+\iy.
\ac
$$

\no {\bf Here and below} {\it for each real harmonic  function $q$
in  $\C_+$ we introduce an analytic function $w=p+iq$  in $\C_+$,
where $(-p)$ is some harmonic conjugate of $q$ for $\C_+$.} Recall
the needed properties of the functions $q\in\cS\cC$. It is well
known, that $p\in C(\ol\C_+)$ and ${1\/2\pi}\D q=\m_q$ (in a sense of
distribution) is a so-called Riesz measure of the function $q$.
Moreover,
\[
\lb{T2.2.1} \pi\m_q((\x_1,\x_2))=p(\x_2)-p(\x_1), \ \ \ \ {\rm for \
any }\ \ \x_1<\x_2,\  \x_1,\x_2\in \R.
\]
Below we need the well-known identity
\[
\lb{id.1} {\pa q(\l)\/\pa \n}=\n\int_{\R}{d\m_q(t)\/(t-\x)^2+\n^2},\ \
\ \l=\x+i\n\in\C_+,
\]
which yields ${\pa q(z)\/\pa \n}\ge 0, \l\in\C_+$. Moreover,
$q(\x)=q(\x\pm i0), \x\in \R.$ It is well known that if $q\in
\cS\cC$, then $\int_{\R}{d\m_q(t)\/1+t^2}+
\int_{\R}{q_-(t)dt\/1+t^2}<+\iy$ and
\[
\lb{L2.7.1} a=\lim\sup_{\l\to\iy}{q(\l)\/|\l|}=
\lim_{\n\to+\iy}{w(i\n)\/i\n}=\lim_{\n\to+\iy}{q(i\n)\/\n}=\lim_{\x\to
\pm \iy}{p(\x)\/\x}\ge 0,
\]
 Define a new classes
$\cS\cK_p$.

\no {\bf Definition}. {\it A real function $q\in \cS\cC$ belongs
to the class $\cS\cK_n, n\in\Z_+:=\Z\cap [0,\iy)$ if
$Q_0^{\pm}+Q_{2n}^{\pm}<+\iy$ and
$$
q(i\n)=\n+{Q_0\/\n}-{Q_2\/\n^{3}}+\dots+(-1)^n{Q_{2n}+o(1))\/\n^{2n+1}}\qq  \as
\qq \n\to+\iy.
$$
 The function $w=p+iq$
in $\C_+$ is defined by $ 
w(\l)=\l+{1\/\pi}\int_{\R}{q(t)dt\/t-\l},\ \ \ \ \l\in\C_+$. }


 Now we recall the well known Nevanlinna Theorem  (see \cite{A65}).

\begin{theorem} \lb{TNe}
  \no i) Let $\m$ be a Borel measure on
$\R$ such that  $\int_{\R}(1+\x^{2m})d\m(\x)<+\iy$ for some
integer $m\ge 0$. Then for each $r>0$ the following asymptotics holds true
as $\l \to\iy, \l\in \mD_r$:
$$
\int_{\R}{d\m(t)\/t-\l}=
-\sum_{j=0}^{2m}{Q_j\/\l^{j+1}}+{o(1)\/\l^{2m+1}},\ \qqq
Q_j=\int_{\R}\x^jd\m(\x),\ 0\le j\le 2m.
$$
\no ii) Let $F$ be an analytic function in $\C_+$ such that $\Im
F(\l)\ge 0,\ \l\in\C_+$ and
$$ \Im
F(i\n)={c_0\/\n}\!+\!\dots\!+\!{c_{2m-1}\/\n^{2m}}\!+\!{O(1)\/\n^{2m+1}}\qqq
\as \qq \n\!\to\!\iy
$$
 for some $c_0,..c_{2m-1}, m\ge 0$.  Then $
F(\l)=C+\int_{\R}{d\m(t)\/t-\l},\ \l\in\C_+$, for some Borel measure
$\m$ on $\R$ such that $\int_{\R}(1+t^{2m})d\m(t)<\iy$ and $C\in
\R$.
\end{theorem}

Let $\dk:\C_+\to \K(h)$ be a conformal mapping for some $h\in
C_{us}$. Assume that $\dk$ has asymptotics
\[
\lb{askm}
\dk(\l)=\l+\sum_{j=0}^{2m}{c_j\/\l^{j+1}}+{o(1)\/\l^{2m+1}} \qqq \as
\qq \Im \l\!\to\!\iy
\]
for some $c_0,..c_{2m}, m\ge 0$. We introduce the
function $k_{(m)}$ and the Dirichlet integral $I_m$  by
$$
\begin{aligned}
& k_{(m)}(\l)={1\/\pi}\int_{\R}{t^m\gq(t)\/t-\l}dt=\l^m(\dk(\l)-\l)+G_m(\l),\
\ \ \ \l\in\C_+,
\\
& G_m(\l)=\sum_{1}^{m}Q_{j-1}\l^{m-j},\qq
Q_j={1\/\pi}\int_{\R}t^j\gq(t)dt,\qq I_m={1\/\pi}\iint_{\C_+}|k^{'}_{(m)}(\l)|^2d\x d\n.
\end{aligned}
$$

\begin{theorem}  \lb{T1.7}
Let $\dk:\C_+\to \K(h)$ be a conformal mapping for some $h\in
C_{us}$ such that asymptotics \er{askm} at $m=1$   and
$Q_0+Q_{2}<\iy$ hold true. Then $I_0+I_1<\iy$ and the following identity holds
true:
\[
\lb{T1.7.2} I_1+P_{2}=Q_{2}+{Q_{0}^2\/2},\qq \where 
P_2={1\/\pi}\int_{\R}t^2\gq(t)d\gp(t).
\]

\end{theorem}

\subsection{Conformal mapping  for $v\in \mH$}
Introduce a simple function
\[
\lb{dx} \e(z):=z+\sqrt{z^2-1},\ \ \ \ z\in \C\sm [-1,1],\ \ {\rm
and}\ \ \ \  \e(z)=2z+O(1/z)\qq as \qq |z|\to \iy.
\]
which is a  conformal mapping from $\C\sm [-1,1] $ onto $\C\sm
\dD_1$. Note that $\e(z)=\ol \e(\ol z), z\in \C\sm [-1,1]$ since
$\e(z)>1$ for any $z>1$. Due to the properties of the Lyapunov
functions we have $|\e(\D_j(\z))|>1, \z\in \cR_+=\{\z\in\cR: \Im
\z>0\}$. Thus we can introduce the quasimomentum $k_j, j\in \N_3$
(we fix some branch of $\arccos$ and $\D_j(z)$) and the function
$q_j$ by
\[
\lb{dkj1} k_j(z)=\arccos \D_j(z)=i\log \e(\D_j(z)),\ \ q_j(z)=\Im
k_j(z)=\log |\e(\D_j(z))|,\ \
\]
$z\in \mR_0^+=\C_+\sm \b_+, \b_+=\!\bigcup_{\b\in \cB_\D\cap \C_+}
[\b, \b+i\iy)$ where $\cB_\D$ is the set of all branch points of the
function $\D$. The branch points of $k_j$ in $\C_+$ belong to
$\cB_\D$. Define the {\bf averaged quasimomentum} $\dk$ and real
functions $\gp, \gq$ by
\[
\dk(z)=\gp(z)+i\gq(z)={1\/3}\sum_1^3 k_j(z), \ \ \  q(z)=\Im k(z), \
\ z\in \mR_0^+.
\]

The Lyapunov function $\D(\z)$ is analytic on some $3$--sheeted
Riemann surface $\cR$ and $q(\z)=\log |\e(\D(\z))|$ is the
single-valued function on $\cR_+$ and is the imaginary part of the
(in general, many-valued on $\cR_+$) quasimomentum $k(\z)$ given by
$$
k(\z)=p(\z)+iq(\z)=\arccos\D(\z)=i\log\e(\D(\z)).
$$
We denote by $q_j(\l)$, $(\l,j)\in\C_+\ts \N_3$, the branches of
$q(\z)$ and by $p_j(\l)$, $k_j(\l)$, $\l\in\cR_+^o$, the
single-valued branches of $p(\z)$, $k(\z)$, respectively. Let
$\l=\x+i\n\in\C$ be the natural projection of $\z\in\cR$ given by
$\dP(\z)=\l$. We denote by $q_j(\l)$, $(\l,j)\in\C_+\ts \N_3$, the
branches of $q(\z)$ and by $p_j(\l)$, $k_j(\l)$, $\l\in\cR_+^o$, the
single-valued branches of $p(\z)$, $k(\z)$, respectively. Let
$\l=\m+i\x\in\C$ be the natural projection of $\z\in\cR$ given by
$\dP(\z)=\l$. Recall that an AQ $\dk(\l)={1\/3}\sum_1^3 k_j(\l)$ is
a conformal mapping from $\C_+$ onto some $\K(h)$.
 We describe properties of Lyapunov functions $\D$ on
Riemann surfaces.

\begin{theorem}
\lb{T41} Let $\z\to \l=\gp(\z)\in \C,\z\in \cR$ be the natural
projection. Then the function $q(\z)=\log |\e(\D(\z))|$ is
subharmonic on the Riemann surface $\cR$ and has the following
asymptotics:
\[
\lb{41r}
 q(\z)=\x+O(|\l|^{-{1\/2}})\qq \as \qq |\l|\to \iy, \  \z\in
\ol\cR_+.
\]
Let, in addition,  $\g_o=(\l^-, \l^-)$ be some open gap in the
spectrum $\gS_3$ and let an interval $\o\ss \g_o$, and $\D_3(\o)\ss
[-1,1]$ for some labeling.

\no i) If $\D_j(\o)\ss \R\sm [-1,1]$ for some $j\in \N_3$, then the
Riemann surface $\cR$ is 2-sheeted.
\\
 ii) Let $\D_1(\l)$ be complex  on $\o$. Then  the Riemann surface $\cR$
is 3-sheeted; the functions $\D_1(\l)$ and $k_{1}+k_2$ have analytic
extensions from $\cR_+^o$ into $\cR_{\o}=\cR_+^o\cup \cR_-^o\cup \o$
such that
\[
\lb{45a4} \D_1(\l)=
\ca \D_1(\l)\ \ \ &if\ \ \ \ \l\in \cR_+^o,\\
       \ol\D_{2}(\ol \l)\ \ \ &if\ \ \ \ \l\in \cR_-^o,\ac
\]
\[
\lb{464}
\ca \ p_1(\l)+p_{2}(\l)={\rm const}\in 2\pi \Z,
 \\
\  0<q_1(\l)=q_{2}(\l)\le \|v\|,\ \ \ \ \ac
\qqq \l\in \g_o,
\]
\[
\lb{474}
 q_1(\l)+q_{2}(\l)=q_1(\ol \l)+q_{2}(\ol \l)>0,\ \
\ \l\in \cR_{\g_o}.
\]
\end{theorem}

 In Theorem \ref{T41}, ii) we have
$\D_2=\ol\D_1$ on $\g_o$. Thus we have 3 different functions $\D_j,
j\in \N_3$ and then the Riemann surface $\cR$ is 3-sheeted.

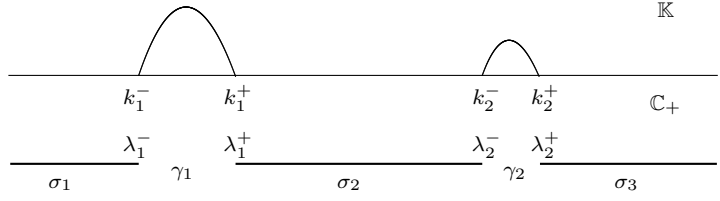
\begin{figure}
\tiny
\unitlength 0.7mm 
\linethickness{0.4pt}
\ifx\plotpoint\undefined\newsavebox{\plotpoint}\fi 
\begin{picture}(138.675,40.737)(0,0)
\put(5.325,26.75){\line(1,0){133.35}}
\qbezier(29.825,26.75)(38.575,52.737)(48.025,26.575)
\qbezier(94.4,26.75)(99.212,40.138)(105.075,26.575) \thicklines
\put(5.675,10.){\line(1,0){24.15}}
\put(48.025,10.){\line(1,0){46.375}}
\put(105.25,10.){\line(1,0){33.25}}
\put(128.875,39.175){\makebox(0,0)[cc]{$\K$}}
\put(128.875,21.175){\makebox(0,0)[cc]{$\C_+$}}
\put(29.65,22.55){\makebox(0,0)[cc]{$k_1^-$}}
\put(48.55,22.55){\makebox(0,0)[cc]{$k_1^+$}}
\put(95.275,22.55){\makebox(0,0)[cc]{$k_2^-$}}
\put(106.3,22.55){\makebox(0,0)[cc]{$k_2^+$}}
\put(29.65,13.625){\makebox(0,0)[cc]{$\l_1^-$}}
\put(48.55,13.625){\makebox(0,0)[cc]{$\l_1^+$}}
\put(95.275,13.625){\makebox(0,0)[cc]{$\l_2^-$}}
\put(106.3,13.625){\makebox(0,0)[cc]{$\l_2^+$}}
\put(15.125,6.1){\makebox(0,0)[cc]{$\s_1$}}
\put(69.375,6.1){\makebox(0,0)[cc]{$\s_2$}}
\put(121.525,6.1){\makebox(0,0)[cc]{$\s_3$}}
\put(38.05,8.2){\makebox(0,0)[cc]{$\g_1$}}
\put(100.525,8.2){\makebox(0,0)[cc]{$\g_2$}}
\end{picture}
\lb{fig3} \caption{\footnotesize The domain $\K$ and
$k_n^\pm=k(\l_n^\pm)$ }
\end{figure}


\subsection{Conformal mapping  for $v'\in \mH$}

\begin{lemma}  \lb{Taqm}
Let $\l\in \mD_r,  |\l|\to \iy,$ for any fixed $ r>0$, and
\[
T(\l)=e^{-i\dk_m(\l)+o(\l^{-2m})},\qqq
\dk_m(\l)=\l-\sum_0^{2m}a_j\l^{-j},
\]
for some $(v,m, a_j)\in \mH\ts \N\ts \C$. Then
\[
\lb{akm}  \dk(\l)=\l-{\tes{2\/3}}\dk_m(\l)+o(\l^{-2m}),
\]

\end{lemma}

 \no {\bf Proof.} Let $\l\in \mD_r, |\l|\to \iy$  for any $ r>0$.
 From \er{tt4} and \er{a1} we obtain
 $$
\begin{aligned}
\t_3(\l)+O(e^{-\n})=T(\l)=e^{-i\dk_m(\l)+o(\l^{-2m})}.
\end{aligned}
$$
Substituting this into $k_3(\l)=i\log \e(\D_3(\l))$ using \er{dx} we
obtain
\[
\lb{kj1}
\begin{aligned}
k_3(\l)=i\log [\t_3(\l)+O(e^{-\n})]=i\log
[e^{-i\dk_m(\l)+o(\l^{-2m})}+O(e^{-\n})]=\dk_m(\l)+o(\l^{-2m}).
\end{aligned}
\]
Consider asymptotics of $k_1(\l)+k_2(\l)$. From \er{} we have
$\D_j(\l)={1\/2}\t_j^{-1}(\l)+O(e^{-\n}), j=1,2$, which yields
\[
\lb{kj2}
 k_j(\l)=i\log \e(\D_j(\l))=i\log \e(z)=i\log
[\t_j^{-1}(\l)+O(e^{-\n})],
\]
and then
$$
\lb{kj1=3}
\begin{aligned}
& k_1(\l)+k_2(\l)=i\log
[\t_1^{-1}(\l)+O(e^{-\n})][\t_2^{-1}(\l)+O(e^{-\n})]=i\log
[{1\/\t_1(\l)\t_2(\l)}+O(e^{-\n})]
\\
& =i\log [\t_3(\l)e^{-i\l}+O(e^{-\n})]=i\log
[e^{-i\l-i\dk_m(\l)+o(\l^{-2m})}+O(e^{-\n})]
=\l+\dk_3(\l)+o(\l^{-2m}).
\end{aligned}
$$
Collecting all these estimates we obtain \er{akm}. \BBox

\no {\bf Proof of Theorem \ref{T5}.} Due to Theorem \ref{Tk} the AQ
$\dk$ is a conformal mapping from $\C_+$ onto $\K(h)$ for some $h\in
C_{uc}(\R)$ with asymptotics $\dk(\l)=\l+O(1/\l)$ as  $\l\in \mD_r,
|\l|\to \iy$ for any fixed $ r>0$. Then using Lemma \ref{Taqm},
asymptotics of the trace $T(\l)$ from Lemma \ref{Tt2} give
$$
\tes
\dk(\l)=\l-{2\/3\l}\cH_0-{2\/3\l^2}\cH_1-{2\/3\l^3}\cH_2+{o(1)\/\l^3},
$$
 and the Nevanlinna Theorem \ref{TNe} gives
$$
\tes \dk(\l)=\l-{Q_0\/\l} -{Q_1\/\l^2}-{Q_2+o(1)\/\l^3}
$$
as  $\l\in \mD_r, |\l|\to \iy$. This yields \er{aqm1} and
$Q_j={2\/3}\cH_j, j=0,1,2$. The function $\gq(\l)$ has a symmetric
extension from $\C_+$ into $\C$ by $\gq(\l)=\gq(\ol\l), \l\in \C$
and this new function belongs to $\cS\cC$ see \cite{K08}. Then due
to the identity \er{T1.7.2} at $p=1$ we obtain \er{aqm2}.  \BBox

\no {\bf \bf Proof of Theorem \ref{T6}. } Recall that we have gaps
$\g_{\o}=(\l_{\o}^-,\l_{\o}^+), \o=(n,j)\in \O=\Z\ts \N_2$.
Due to  \er{474} we have $\gq(\l)={2\/3}q_1(\l)$ for $\l\in\g_{\o}$.
The  estimate \er{464} gives $\sup_{\l\in\g_{\o} } q_1(\l)\le \|v\|$.
This and the identity $\cH_2={3\/2}Q_2$ from \er{aqm2} imply
\[
\begin{aligned}
\lb{hq1x}
\cH_2={3\/2}Q_2=\sum_{\o\in \O} {3\/2\pi}\int_{\g_{\o}}\l^2\gq(\l)d\l
   \le  \sum_{\o\in \O} {\|v\|\/\pi} \int_{\g_{\o}}\l^2d\l
   \\
=\sum_{\o\in \O} {\|v\|\/3\pi}\sum_{\o\in \O}[(\l_{\o}^+)^3-\l_{\o}^-)^3]\le
\sum_{\o\in \O} {\|v\|\/3\pi}|\g_\o|(|\l_{\o}^+|+|\l_{\o}^-|)^2,
\end{aligned}
\]
which yields \er{H1}. We show \er{H2}.
Using the estimate $\sup_{\l\in\g_{\o} } q_1(\l)\le \big({2|\g_\o|} \t_\bu^4\big)^{1\/6}$ for $\l\in \g_\o$ and \er{rqmb}, \er{hq1x}   we obtain
$$
\begin{aligned}
\lb{hq1}
\cH_2=\sum_{\o\in \O} {3\/2\pi}\int_{\g_{\o}}\l^2\gq(\l)d\l
   \le  \sum_{\o\in \O} {\big({2|\g_\o|} \t_\bu^4\big)^{1\/6}\/\pi} \int_{\g_{\o}}\l^2d
\le {\big(2 \t_\bu^4\big)^{1\/6}\/3\pi}\sum_{\o\in \O} |\g_\o|^{7\/6} ({\l_{\o}^+} +{\l_{\o}^-})^2,
\end{aligned}
$$
which yields \er{H2}.
\BBox

\section {Appendix \lb{Sec7}}
\setcounter{equation}{0}

From the well-known GershgorinTheorem (see Sect. 7.2 [La]) we obtain

\begin{lemma}  \lb{TG}
Let $A, B$ be $N\ts N$-matrices and let $A$ be normal.
Let $\l_j,\l_j^0,j\in \N_N$ be eigenvalues of $A$ and $A+B$.
Then $\l_j=\l_j^0+O(|B|)$ as $B\to 0$ for each $j\in \N_N$.
\end{lemma}

Note that this result follows from the identity $ \det \ma A_1&
A_2\\ A_{3}& A_4\am=\det A_4\det (A_1-A_2A_4^{-1}A_3), $ for the
corresponding matrices $A_1,..,A_4, \det A_4\ne 0$.

\no  Define the domain $G_n=G_n(1)=\{0< t_n<...<t_1< 1\}\ss \R^n$.
Recall results from \cite{K10}.

\begin{lemma} \lb{TA1}
Let $h_1,...,h_n\in L^2(0,1)$ and $u_n=t_1-t_2+\dots -(-1)^nt_n$ for
some $n\ge 3$. Then
\[
\lb{TA-1} \int_0^1dt\int_0^t e^{i2\l(t-s)}h_1(t)h_2(s)ds = {i\/2\l
}\lt(\int_0^1h_1(t)h_2(t)dt+o(1)\rt),
\]
\[
\lb{TA-2} \int_0^1dt\int_0^t
e^{-i2\l(t-s)}h_1(t)h_2(s)ds={o(e^{2\n})\/|\l|},
\]
\[
\lb{71} \int_0^1dt\int_0^t\cos \l (1-2t+2s)h_1(t)h_2(s)ds ={i\cos
z\/2\l }\lt(\int_0^1h_1(t)h_2(t)dt+o(1)\rt),
\]
\[
\lb{TA-4} \int_{G_n} e^{\pm i\l (1-2u_n)}\prod_1^n h_j(t_j)dt
={o(e^{\n})\/|\l|},
\]
as $\l \in \mD_r$ and  $|\l|\to \iy$ for any fixed $r>0$, where
$\n=\Im \l$.
\end{lemma}

\begin{lemma} \lb{TA3z}
Let $h\in L^2(0,1)$ and let $u_n=t_1-t_2+\dots -(-1)^nt_n$ for some
$n\ge 2$. Then
\[
\lb{ab2} \rt|\int_{G_n} e^{\pm i\l (1- 2 u_n)}\prod_1^n h(t_j) dt
\rt|\le {e^{\n}\/(4\n)^{{n\/4}}}{\|h\|^{n}\/n!},\qqq \n=\Im \l,
\]
as $\l \in \mD_r$ and  $|\l|\to \iy$ for any fixed $r>0$.
\end{lemma}

In the next lemma we recall   the asymptotic estimates of the
following integral
$$
\begin{aligned}
F_n^\pm(x)=\int_0^xe^{\pm i\l _nt}p(t)dt,\ \ \ \ G_n^{\pm
}(x)=\int_0^xe^{\mp i\l t}q(t)F_n^{\pm }(t)dt,\ \ \ \
\\
p,q\in L^2(0,2\pi ),\qq (x,\l_n,n)\in [0,2\pi ]\ts \C \ts \Z.
\end{aligned}
$$

For $p\in L^2(0,2\pi )$ we have the Fourier series $ p(x)=\sum
p_je^{ijx}$.

\begin{lemma}
\label{Ta6.1} Let  $C_0=4e^\pi$ and  $|z_n- \pi n|\le {1\/4}$. Then
\[
|F_n^{\pm}(x)|\le C_0u_{\pm n},\ \ \ |G_n^{\pm}(x)|\le C_0(u_{\pm
n}(q)u_{\pm n}(p)+v_{\pm n}^2+ w_{\pm n}^2),
\]
\[
u_{\pm }(p)=(p_{\pm n}(p))_{n\in \Z}, \ \ v_{\pm }=(v_{\pm n})_{n\in
\Z},\qq  w_{\pm }=(w_{\pm n})\in \ell^d(\Z), \qq \forall \ \ d>2,
\]
where
$$
u_{n}(p)= \sum {|p_j|\/ |j \pm n+i|},\ \ v_{n}^2=\sum {|q_jp_{-j}|\/
|j \pm n+i|},\ \ w_{n}^2= \sum {|p_jq_{s}|\/ |j\pm n+i||j+s+i|},\ \
n\in \Z ,
$$

\end{lemma}

\

 In our  proof some Lemma about the zeros of the function we need
 the following simple fact.

\bigskip

\begin{lemma} Let $r\in [0,{\pi\/2}]$ and $|z-\pi n|\ge r$ for all  $n\in \Z$. Then we have
\[
\lb{FuT1} e^{|\Im z|}\le {2\/(1-e^{-2r})}|\sin z|,
\]
and, in particular, if $r={\pi\/4}$, then ${2\/(1-e^{-2r})}\le {8\/3}$.

\end{lemma}

\no {\bf Proof.}  Sufficiently consider $z\in \ol\C_+$. We have
$
2|\sin z|e^{|\Im z|}=2|e^{iz}\sin z|=|1-e^{i2z}|.
$
Then the max principle for the ${1\/ 1-e^{i2z}}$ for the domain
$
\{z\in \ol\C_+: |z-\pi n|\ge r, n\in \Z\}
$
and periodic property give that we have to check Lemma for $|z|=r,
\Im z\ge 0$. Let $w=1-e^{i2z}$. We have
$$
\tes
|z|=r={1\/2}|\log (1-w)|\le {1\/2}\Big(|w|+{|w|^2\/2}+{|w|^3\/3}...
\Big) ={1\/2}\log (1-|w|).
$$
Then $|w|\ge 1-e^{-2r}$, which gives \er{FuT1}. Note that if
$r={\pi\/4}$, then $e^{\pi\/2}>e^{{3\/2}}>(2.7) \cdot \sqrt
{(1.6)^2}=4.32>4$ and $ {1-e^{-\pi/2}\/2}\ge {3\/8} $, which yields
the needed estimate. \BBox

\setlength{\itemsep}{-\parskip} \footnotesize \no  {\bf
Acknowledgments.}   E. K. was supported by the RSF grant  No.
19-71-30002.

\

The author (E.K.) states that there is no conflict of interest.

The author confirms that the data supporting the findings of
this study are available within the article
 and its supplementary materials.


\end{document}